\documentclass[twocolumn]{aastex63} 

\usepackage{url}
\usepackage[colorlinks=true]{hyperref}
\usepackage{graphicx}
\usepackage{overpic}

\newcommand{\G}{\mathcal{G}}

\newcommand{\appropto}{\mathrel{\vcenter{\offinterlineskip\halign{\hfil$##$\cr\propto\cr\noalign{\kern2pt}\sim\cr\noalign{\kern-2pt}}}}}
\newcommand{\bjdtdb}{\ensuremath{\rm {BJD_{TDB}}}}
\newcommand{\feh}{\ensuremath{\left[{\rm Fe}/{\rm H}\right]}}
\newcommand{\teff}{\ensuremath{T_{\rm eff}}}
\newcommand{\logg}{\ensuremath{\,{\rm log}\,{g}}}

\newcommand{\msun}{\ensuremath{\,{\rm M_\Sun}}}
\newcommand{\rsun}{\ensuremath{\,{\rm R_\Sun}}}
\newcommand{\lsun}{\ensuremath{\,{\rm L_\Sun}}}

\newcommand{\mj}{\ensuremath{\,{\rm M_{\rm J}}}}
\newcommand{\rj}{\ensuremath{\,{\rm R_{\rm J}}}}
\newcommand{\degree}{\ensuremath{\,^{\circ}}}

\newcommand{\fave}{\langle F \rangle}
\newcommand{\fluxcgs}{10$^9$ erg s$^{-1}$ cm$^{-2}$}

\begin{document}

\title{Revisiting the Full Sets of Orbital Parameters for the XO-3 System:\\ No evidence for Temporal Variation of the Spin-Orbit Angle}  

\shortauthors{Worku et al.}

\author{Keduse Worku} 
\affil{Department of Astronomy, Yale University, New Haven, CT 06511, USA}

\author[0000-0002-7846-6981]{Songhu Wang}
\affil{Department of Astronomy, Indiana University, Bloomington, IN 47405, USA}

\author[0000-0002-0040-6815]{Jennifer Burt}
\affil{Jet Propulsion Laboratory, California Institute of Technology, 4800 Oak Grove Drive, Pasadena, CA 91109, USA}

\author[0000-0002-7670-670X]{Malena Rice} 
\affil{Department of Astronomy, Yale University, New Haven, CT 06511, USA}

\author[0000-0002-0376-6365]{Xian-Yu Wang} 
\affil{National Astronomical Observatories, Chinese Academy of Sciences, Beijing 100012, China}
\affil{University of Chinese Academy of Sciences, Beijing, 100049, China}

\author[0000-0002-7846-6981]{Yong-Hao Wang}
\affil{School of Physics and Astronomy, Sun Yat-Sen University, Zhuhai, 519082, China}

\author{Steven S. Vogt}
\affil{UCO/Lick Observatory, Department of Astronomy and Astrophysics, University of California at Santa Cruz, Santa Cruz, CA 95064, USA}

\author[0000-0003-1305-3761]{ R. Paul Butler}
\affil{Earth and Planets Laboratory, Carnegie Institution for Science, 5241 Broad Branch Road NW, Washington, DC 20015, USA}

\author[0000-0003-3216-0626]{Brett Addison}
\affil{University of Southern Queensland, Centre for Astrophysics, USQ Toowoomba, West Street, QLD 4350 Australia}

\author{Brad Holden}
\affil{UCO/Lick Observatory, Department of Astronomy and Astrophysics, University of California at Santa Cruz, Santa Cruz, CA 95064, USA}

\author{Xi-Yan Peng}
\affil{Shanghai Astronomical Observatory, Chinese Academy of Sciences, Shanghai 200030, China}

\author[0000-0001-8037-1984]{Zhen-Yu Wu}
\affil{National Astronomical Observatories, Chinese Academy of Sciences, Beijing 100012, China}
\affil{University of Chinese Academy of Sciences, Beijing, 100049, China}

\author{Xu Zhou}
\affil{National Astronomical Observatories, Chinese Academy of Sciences, Beijing 100012, China}

\author[0000-0001-5162-1753]{Hui-Gen Liu}
\affil{School of Astronomy and Space Science and Key Laboratory of Modern Astronomy and Astrophysics in Ministry of Education, Nanjing University, Nanjing 210093, China}

\author[0000-0003-3491-6394]{Hui Zhang}
\affil{Shanghai Astronomical Observatory, Chinese Academy of Sciences, Shanghai 200030, China}

\author[0000-0003-1680-2940]{Ji-Lin Zhou}
\affil{School of Astronomy and Space Science and Key Laboratory of Modern Astronomy and Astrophysics in Ministry of Education, Nanjing University, Nanjing 210093, China}

\author[0000-0002-3253-2621]{Gregory Laughlin}
\affil{Department of Astronomy, Yale University, New Haven, CT 06511, USA}

\newcommand{\Nice}{\textit{Nice}}

\begin{abstract} 
We present 12 new transit light curves and 16 new out-of-transit radial velocity measurements for the XO-3 system. By modelling our newly collected measurements together with archival photometric and Doppler velocimetric data, we confirmed the unusual configuration of the XO-3 system, which contains a massive planet ($M_{\rm P}=11.92^{+0.59}_{-0.63}{\,\rm \mj}$) on a relatively eccentric ($e=0.2853^{+0.0027}_{-0.0026}$) and short-period ($3.19152 \pm 0.00145\,$day) orbit around a massive star ($M_*=1.219^{+0.090}_{-0.095} M_{\odot}$). Furthermore, we find no strong evidence for a temporal change of either $V\sin i_{*}$ (and by extension, the stellar spin vector of XO-3), or the transit profile (and thus orbital angular momentum vector of XO-3b). We conclude that the discrepancy in previous Rossiter-McLaughlin measurements ($70.0\degree \pm 15.0\degree$ \citep{Hebrard2008}; $37.3\degree \pm 3.7\degree$ \citep{Winn2009}; $37.3\degree\pm 3.0\degree$ \citep{Hirano2011}) may have stemmed from systematic noise sources.
\end{abstract}

\section{Introduction} \label{sect1}

The existence of hot Jupiters, giant planets orbiting perilously close to their parent stars, was wholly unpredicted; as a consequence, their initial discoveries twenty-five years ago occurred with very high signal-to-noise. In retrospect, this historical development was largely a consequence of the emptiness of the inner reaches of the Solar System. 

Although it has been suggested that hot Jupiters may form \textit{in situ} \citep{Batygin2016}, conventional wisdom still holds that they form at larger distances -- where cold, ice-based materials are plentiful -- and then migrate inward \citep{Bodenheimer2000}. Over the past two decades, two distinct and competing long-distance migration mechanisms have been established: namely, quiescent disk migration \citep{Lin1996} and violent dynamical migration (Lidov-Kozai cycling with tidal friction \citep{Wu2007, Fabrycky2007, Naoz2016}; planet-planet scattering \citep{Rasio1996, Nagasawa2008}; or secular interactions \citep{Wu2011, Petrovich2015}). Nonetheless, the most workable process for delivering a ``normal Jupiter'' to its final location remains controversial.

Measurements of stellar obliquity (i.e., the sky-projected angle, $\lambda$, between the orbital angular momentum vector of a transiting planet and its host star's spin vector) through the Rossiter-Mclaughlin effect \citep[R-M effect;][]{Rossiter1924, McLaughlin1924, Queloz2000} were initially thought to provide a zeroth-order discriminating test between quiescent disk-driven migration and violent dynamical migration. Hot Jupiters with low spin-orbit angles were thought to have migrated through the disk, while those with high spin-orbit angles were believed to owe their orbits to high-eccentricity dynamical migration. 

The origin and evolution of spin-orbit misalignment has since been extensively studied (see \citealt{Winn2015} and references therein), however, and the connection between the spin-orbit misalignment and the migration process may be more complicated than was initially thought. These misalignments can either be primordial (chaotic star formation \citep{Bate2010, Fielding2015}; magnetic star-disk interactions \citep{Lai2011, Spalding2014}; torques from the stellar companions \citep{Batygin2012}), with planets born in tilted disks, or they can be modified later by post-migration evolution (tidal and magnetic realignment \citep{Winn2010, Dawson2014, Li2016}; gravitational perturbation from the companions \citep{Innanen1997, Li2014, Storch2014, Lai2016, Gomes2017}; internal gravity waves \citep{Rogers2012}). The spin-orbit angle generally evolves on very long timescales. Under certain circumstances, however, it can vary on an observable timescale (See \citealt{Rogers2012}, for example). 

In this light, the XO-3 system \citep{Johns-Krull2008} has a special importance as one of few transiting planet systems that displays a discrepancy between multiple R-M measurements ($70.0\degree \pm 15.0\degree$, \citealt{Hebrard2008}; $37.3\degree \pm 3.7\degree$, \citealt{Winn2009}; $37.3\degree\pm 3.0\degree$, \citealt{Hirano2011}). This discrepancy, in combination with the planet's unusual mass ($M_{\rm P}=11.7\,M_{\rm J}$, \citealt{Bonomo2017}) -- lying just at the mass limit between giant planets and low-mass stars -- its eccentricity ($e=0.28$, \citealt{Wong2014}), and its short period ($P=3.19\,{\rm day}$, \citealt{Winn2008}), suggests that XO-3 merits further scrutiny.
 
In this paper, we present new transit light curves and Doppler velocimetric measurements for XO-3 to address the source of the discrepancy in previous R-M measurements, as well as implications for the origins of hot Jupiters.
 
The paper is organized as follows. 
\S \ref{sect2} presents 33 photometric transit observations from the literature and 12 new transits, as well as 16 new Doppler velocity measurements of XO-3 that were used in this study. 
\S \ref{sect3} focuses on the characterization of the stellar atmospheric parameters using the APF data. 
\S \ref{sect4} describes the joint analysis of the in-transit photometric and out-of-transit radial velocity measurements. 
\S \ref{sect5} compares these findings with previous results and outlines the potential implications of this study.

\section{Observation and Data Reduction} \label{sect2}

\subsection{Photometry}

Our new photometric dataset, comprised of twelve light curves, was collected using the Xinglong Schmidt and Xinglong $60\,{\rm cm}$ telescopes operated by the National Astronomical Observatories of China (NAOC). These observations span roughly four years, from 2014 March to 2017 December. 

Four of these light curves were obtained using the Xinglong Schmidt telescope \citep{zhou1999, zhou2001}, which utilizes a $4{\rm K} \times 4{\rm K}$ CCD. This CCD has a field of view (FOV) of $94' \times 94'$ and a pixel scale of $1.38''\,{\rm pixel ^{-1}}$. To reduce the initial readout time ($93\,{\rm s}$), we windowed the frames down to $512 \times 512$ pixels, which results in a readout time of $12\,{\rm s}$. A Johnson/Cousins $R$-band filter was used during these observations. 

The remaining eight light curves were obtained with the Xinglong $60\,{\rm cm}$ telescope. The observation conducted on UT 2014-03-02 used a $512 \times 512$ CCD, giving an FOV of $17' \times 17'$, a pixel scale of $1.95''\,{\rm pixel ^{-1}}$, and a readout time of $3\,{\rm s}$. The observations conducted on UT 2015-02-16, UT 2017-11-11, and UT 2017-11-14 used a $1{\rm K} \times 1{\rm K}$ CCD, giving an FOV of $17' \times 17'$, a pixel scale of $0.99''\,{\rm pixel ^{-1}}$, and a readout time of $23\,{\rm s}$. The observations conducted on UT 2016-01-14, UT 2016-02-18, UT 2016-03-05 used a $2{\rm K} \times 2{\rm K}$ CCD, giving an FOV of $36' \times 36'$, a pixel scale of $1.06''\,{\rm pixel ^{-1}}$, and a readout time of $6\,{\rm s}$. All of the observations for this telescope utilized a Johnson/Cousins $R$-band filter, except the one from UT 2017-11-11 which alternated between the Johnson/Cousins $B$-band and $V$-band filters.

As XO-3 is bright (Vmag$=9.86$), we defocused the telescopes to avoid non-linear effects on the CCD. The defocusing method usually requires longer exposure times, which is helpful for increasing the duty cycle of our observations and reducing the scintillation and Poisson noise \citep{Southworth2009}. The focus was kept unchanged during our observation. The exposure time was changed only if required by weather conditions. The telescope time was synchronized with online GPS time servers. The beginning time of each exposure was recorded in the frame header using the UTC time standard, and it was then converted to ${\rm BJD_{\rm TDB}}$ as described in \citet{eastman2010}. All times reported from previous works have been converted to ${\rm BJD_{\rm TDB}}$ for congruency with our time standard.

\begin{figure}[t]
\centering
\includegraphics[width=1\columnwidth]{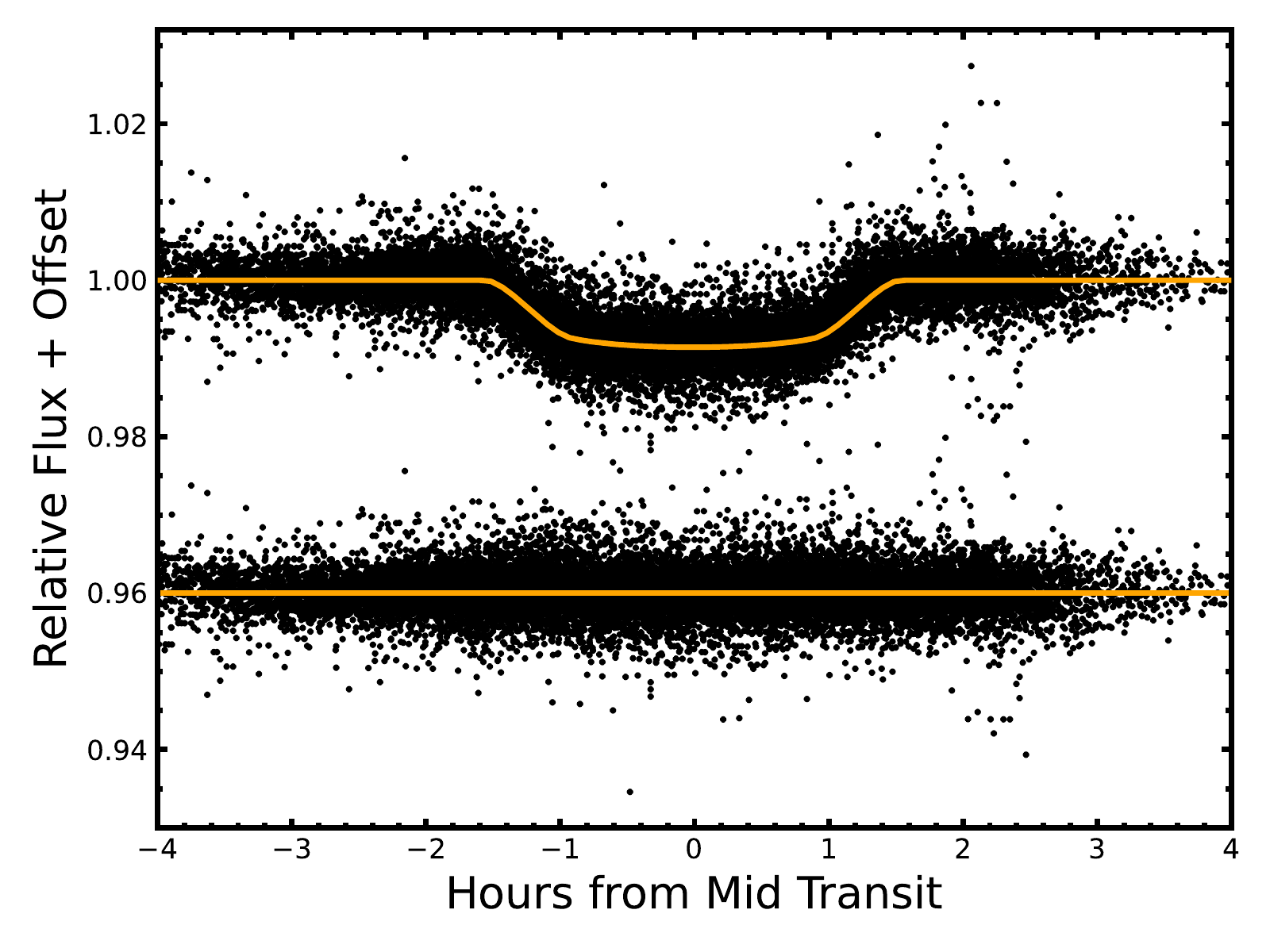}
\caption{Phased light curve of XO-3b including 33 literature transits provided by previous works \citep{Johns-Krull2008, Winn2008, Winn2009, Bonomo2017} and 12 new transits collected in this work.  To estimate the system parameters, these light curves were simultaneously fitted with the radial velocity measurements (Figure~\ref{f3}) as described in detail in \S \ref{sect4}. The orange solid line represents the best-fitting model to all of the data and the residuals are plotted below.}
\label{f1}
\end{figure}

\begin{figure}[t]
\centering
\includegraphics[width=1\columnwidth]{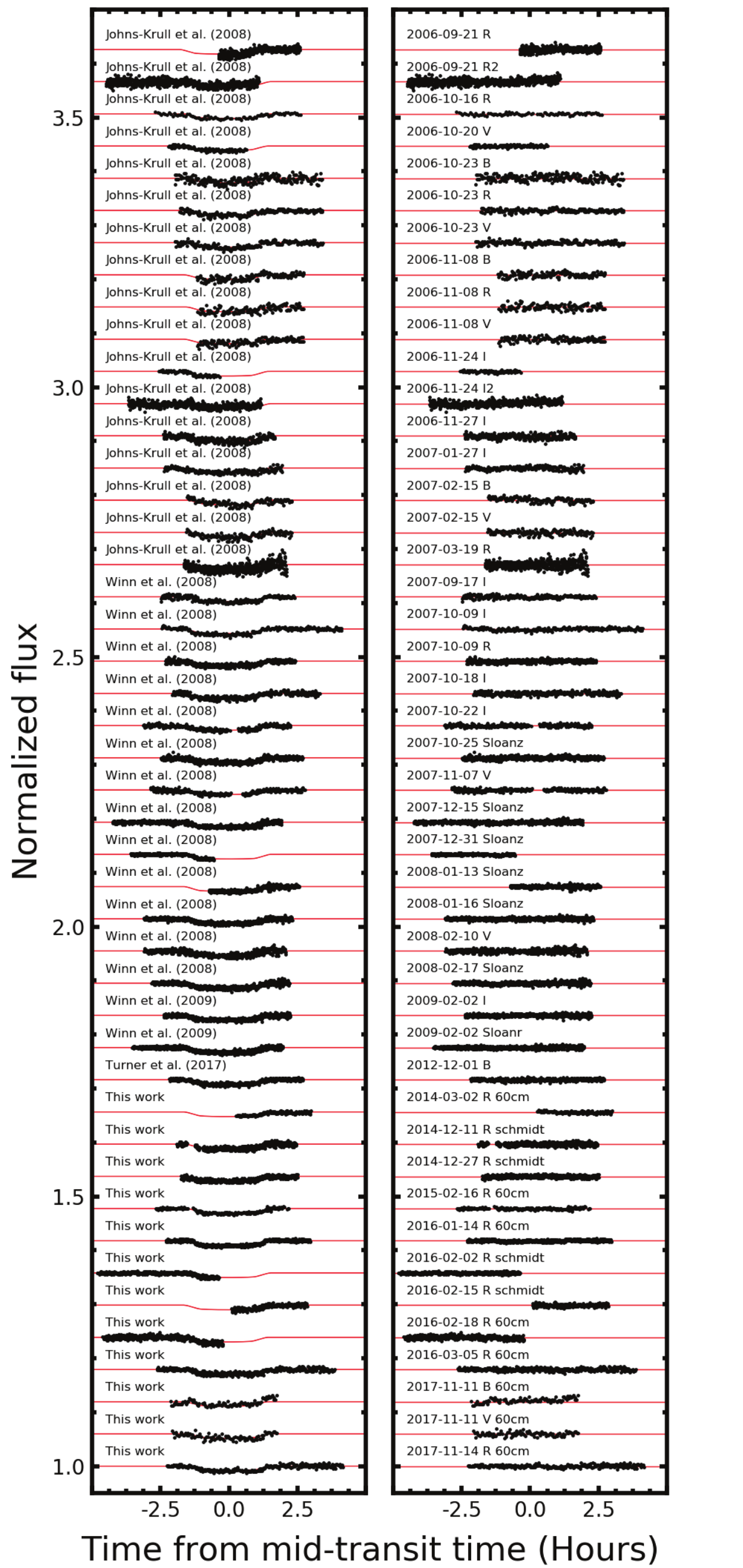}
\caption{Light curves of XO-3b transits, obtained from the literature and as part of this work. The best-fitting models are overplotted as solid red lines in the left panel. The residuals are displayed in the right panel. Both light curves and residuals are sorted by observation date for clarity.}
\label{f2}
\end{figure}

We conducted standard bias and flat-field corrections on all the frames following the procedures described in \citet{WangX2018, WangX2021a, Wang2018a, Wang2018c, WangY2017, WangY2019}. We then performed aperture photometry using SExtractor \citep{Bertin1996}. We identified the best aperture for both the target and reference stars as the one that minimized the root mean square (RMS) of the final differential light curves, which are obtained by comparing XO-3 with three reference stars in the field. Highly discrepant points and/or linear trends presented in these light curves were removed. A summary of the observations and the data reduction procedures are listed in Table~\ref{tab:obs}. The final light curves are presented in Table~\ref{tab:lcs} and plotted in Figures \ref{f1} and \ref{f2}.

\setlength{\tabcolsep}{1.1pt}
\begin{deluxetable*}{ccccccccccc}[!]
\tabletypesize{\scriptsize}
\tablewidth{0pt}
\tablecaption{Overview of Observations and Data Reduction for XO-3}
\tablehead{
\colhead{Date}  &  \colhead{Time}   &   \colhead{Telescope}& \colhead{Band} &  \colhead{Frames}  &  \colhead{Exposure}  & \colhead{Read} & \colhead{Airmass} &\colhead{Moon}    & \colhead{Aperture}\tablenotemark{a}& \colhead{Scatter}\tablenotemark{b}\\
\colhead{(UTC)} &\colhead{(UTC)} & \colhead{} & \colhead{}  &  \colhead{} & \colhead{(second)} & \colhead{(second)} & \colhead{} & \colhead{illum.} & \colhead{(pixels)}  &  \colhead{(mmag)} }
\startdata 
2014 Mar 02 & { }{ }$10:43:00\rightarrow16:01:40$ &  { } Xinglong $60\,{\rm cm}$    &   $R$ &     306  & 60         &3&     $1.06\rightarrow1.88$                      &0.03     &18   &1.7 \\  
2014 Dec 11 & { }{ }$12:05:58\rightarrow16:27:51$ &  { } Xinglong Schmidt           &   $R$ &     464  & 20         &12&    $1.21\rightarrow1.05\rightarrow1.07$       &0.75     &16   &2.5 \\  
2014 Dec 27 & { }{ }$11:15:39\rightarrow15:32:22$ &  { } Xinglong Schmidt           &   $R$ &     515  & 18         &12&    $1.18\rightarrow1.05\rightarrow1.08$       &0.38     &17   &2.2 \\  
2015 Feb 16 & { }{ }$10:37:34\rightarrow16:45:33$ &  { } Xinglong $60\,{\rm cm}$    &   $R$ &     269  & 60         &23&    $1.05\rightarrow1.82$                      &0.08     &30   &1.6 \\  
2016 Jan 14 & { }{ }$10:15:34\rightarrow15:31:46$ &  { } Xinglong $60\,{\rm cm}$    &   $R$ &     336  & 10-25      &6&     $1.17\rightarrow1.05\rightarrow1.15$       &0.25     &23   &1.7 \\  
2016 Feb 02 & { }{ }$11:17:33\rightarrow15:47:20$ &  { } Xinglong Schmidt           &   $R$ &     460  & 25         &12&    $1.05\rightarrow1.35$                      &0.35     &18   &1.5 \\  
2016 Feb 15 & { }{ }$10:43:10\rightarrow13:28:28$ &  { } Xinglong Schmidt           &   $R$ &     198  & 25-55      &12&    $1.05\rightarrow1.15$                      &0.54     &18   &2.4 \\  
2016 Feb 18 & { }{ }$10:35:44\rightarrow14:54:19$ &  { } Xinglong $60\,{\rm cm}$    &   $R$ &     550  & 10-25      &6&     $1.05\rightarrow1.38$                      &0.84     &18   &2.7 \\  
2016 Mar 05 & { }{ }$11:33:32\rightarrow16:02:43$ &  { } Xinglong $60\,{\rm cm}$    &   $R$ &     506  & 25-45      &6&     $1.11\rightarrow2.00$                      &0.15     &20   &2.2 \\  
2017 Nov 11 & { }{ }$11:07:08\rightarrow14:59:15$ &  { } Xinglong $60\,{\rm cm}$    &   $B$ &     97   & 20-30      &23&    $1.82\rightarrow1.12$                      &0.41     &23   &4.8 \\  
2017 Nov 11 & { }{ }$11:08:05\rightarrow14:57:34$ &  { } Xinglong $60\,{\rm cm}$    &   $V$ &     96   & 20-30      &23&    $1.82\rightarrow1.12$                      &0.41     &23   &3.6 \\  
2017 Nov 14 & { }{ }$15:31:09\rightarrow21:57:23$ &  { } Xinglong $60\,{\rm cm}$    &   $R$ &     336  & 20         &23&    $1.08\rightarrow1.05\rightarrow1.55$       &0.14     &25   &2.3     \\
\enddata
\tablenotetext{a}{This column indicates the aperture diameter used in SExtractor.}
\tablenotetext{b}{This column presents the RMS scatter of residuals from the best-fitting model.}
\label{tab:obs}
\end{deluxetable*}

\begin{deluxetable}{ccccc}
\tablewidth{0pt}
\tablecaption{Photometry for XO-3}
\tablehead{
\colhead{${\rm BJD}$} & \colhead{Flux}                 & \colhead{$\sigma_{\rm Flux}$} & \colhead{Instrument} & \colhead{Filter} }
\startdata
  2456719.055590& {}       0.9950&  {}       0.0017&{} Xinglong $60\,{\rm cm}$  &  $R$ \\
  2456719.056318&  {}      0.9943&  {}       0.0017&{} Xinglong $60\,{\rm cm}$  &  $R$ \\
  2456719.057046&    {}    0.9948&  {}       0.0017&{} Xinglong $60\,{\rm cm}$  &  $R$ \\
  2456719.057773&      {}  0.9965&   {}      0.0017&{} Xinglong $60\,{\rm cm}$  &  $R$ \\
  2456719.058501&       {} 0.9956&  {}       0.0017&{} Xinglong $60\,{\rm cm}$  &  $R$ \\
  2456719.059229&    {}    0.9940&  {}       0.0017&{} Xinglong $60\,{\rm cm}$  &  $R$ \\
  2456719.059957&    {}     0.9947& {}        0.0017&{} Xinglong $60\,{\rm cm}$  &  $R$ \\
  2456719.060685&    {}     0.9935&  {}       0.0017&{} Xinglong $60\,{\rm cm}$  &  $R$ \\
  2456719.061412&    {}     0.9962& {}        0.0017&{} Xinglong $60\,{\rm cm}$  &  $R$ \\
  2456719.062140&   {}      0.9943& {}        0.0017&{} Xinglong $60\,{\rm cm}$  &  $R$ \\
  2456719.062868&  {}       0.9944&  {}       0.0017&{} Xinglong $60\,{\rm cm}$  &  $R$ \\
  2456719.063597&  {}       0.9935&  {}       0.0017&{} Xinglong $60\,{\rm cm}$  &  $R$ \\
  ...             &  {}      ... &  {}       ... &{} ...  &  ...
\enddata
\tablecomments{The complete table is available in the machine readable format. We put a portion here just for guidance concerning the form and content.}
\label{tab:lcs}
\end{deluxetable}

\subsection{Velocimetry}

XO-3 has been the subject of a variety of radial velocity (RV) observing campaigns over the past two decades. In this work, we combine all previously published RV data sets with a new set of velocity measurements obtained using the Automated Planet Finder (APF) telescope located at Lick Observatory. The APF couples a 2.4m primary mirror with the slit-fed, Levy echelle spectrograph, which works at a typical spectral resolution of R $\sim$ 110,000 and delivers a peak overall system efficiency (fraction of photons striking the telescope primary that are detected by the CCD) of 15\% \citep{Vogt2014}. The telescope was designed to search for planets in the liquid water habitable zone of nearby stars. The APF is driven by a dynamic scheduling software system that can make minute-to-minute decisions on what target to observe based on the ambient atmospheric transparency, atmospheric seeing, and lunar phase \citep{Burt2015}. This allows the telescope to operate efficiently throughout the year without the need for human supervision.

Like its predecessors on the Keck and Magellan telescopes (HIRES: \citealt{Vogt1994}, and PFS: \citealt{Crane2010}, respectively), the APF uses a gaseous I$_{2}$ cell to imprint a forest of narrow absorption lines on the stellar spectrum before its incidence on the spectrograph slit \citep{Butler1996}. These I$_{2}$ absorption lines create a stable wavelength calibration source and permit the measurement of the spectrometer's point spread function (PSF). For each stellar spectrum, the 5000 - 6200\AA$\,$ region (which contains the highest density of I$_{2}$ lines) is subdivided into $\sim$700 individual 2\AA$\,$ segments, with each segment providing an independent measure of the wavelength, the PSF, and the Doppler shift. Our reported overall stellar velocity from a given spectrum is a weighted mean of the individual segments' velocity measurements. The uncertainty for each velocity is the RMS of the individual segment velocity values about the mean divided by the square root of the number of segments. This “internal” uncertainty primarily represents errors in the fitting process, which are dominated by Poisson statistics. The velocities are expressed relative to the solar system barycenter, but are not referenced to any absolute fiducial point. Since it began scientific operations in Q2 2013, the APF has contributed to a number of planet detections (e.g., \citealt{Burt2014,Burt2021, Fulton2015, Vogt2015, Christiansen2017}) and has showcased its ability to reach internal precisions of $\sim$ 1m/s on bright, quiet stars. Indeed, the APF has consistently achieved internal velocity precision of order $\sigma \lesssim$ 2 m/s on bright (e.g. V $\lesssim$ 8) stars \citep{Vogt2015}. 

Table \ref{tab:rvs} presents our newly collected RV measurements for XO-3b, with 16 individual exposures. The median internal uncertainty for our observations is $\sigma_{i} \approx$ 15.9 m/s with an exposure time of $45\,{\rm mins}$. These large internal uncertainties are driven by (1) the star's high rotational velocity ($V\sin i_{*}=17.3 \pm 0.9{\rm \,km\,s^{-1}}$, see Section~\ref{sect3} for more details), which rotationally broadens the stellar absorption lines, thereby reducing their Doppler information content, and (2) by the star's high effective temperature ($T_{\rm eff}$ = 6471$^{+83}_{-82}$K), which reduces the overall number of absorption lines \citep{Torres2012, Bouchy2011, Beatty2015}.

\begin{deluxetable}{cccccc}
\tablecaption{\label{tab:rvs} APF RVs for XO-3}
\tablehead{
\colhead{${\rm BJD}$} &  &\colhead{RV}          &   \colhead{}       & \colhead{$\sigma_{\rm RV}$} \\
\colhead{}       &  \colhead{}     & \colhead{m s$^{-1}$}      &  \colhead{}    & \colhead{m s$^{-1}$}             
}
\startdata
 2457683.767  &&  -918.53333   &&  14.410   \\ 
 2457683.798  &&  -951.22600   &&  15.857   \\ 
 2457683.830  &&  -946.57400   &&  15.773   \\ 
 2457683.861  &&  -951.54200   &&  14.680   \\ 
 2457683.892  &&  -1031.46600  &&  15.292   \\ 
 2457683.923  &&  -994.45200   &&  14.566   \\ 
 2457683.955  &&  -1016.47200  &&  15.046   \\ 
 2457683.986  &&  -986.96500   &&  14.303   \\ 
 2457684.017  &&  -1077.11600  &&  15.232   \\ 
 2457684.048  &&  -1050.91000  &&  16.088   \\ 
 2457809.638  &&  383.11333    &&  25.398   \\ 
 2457809.669  &&  422.80429    &&  23.685   \\ 
 2457809.700  &&  596.86000    &&  23.145   \\ 
 2457809.732  &&  720.67714    &&  18.881   \\ 
 2457809.763  &&  824.05000    &&  18.534   \\ 
 2457809.794  &&  852.69000    & & 22.612    \\
\enddata
\end{deluxetable}

\begin{figure}[t]
\centering
\includegraphics[width=1\columnwidth,]{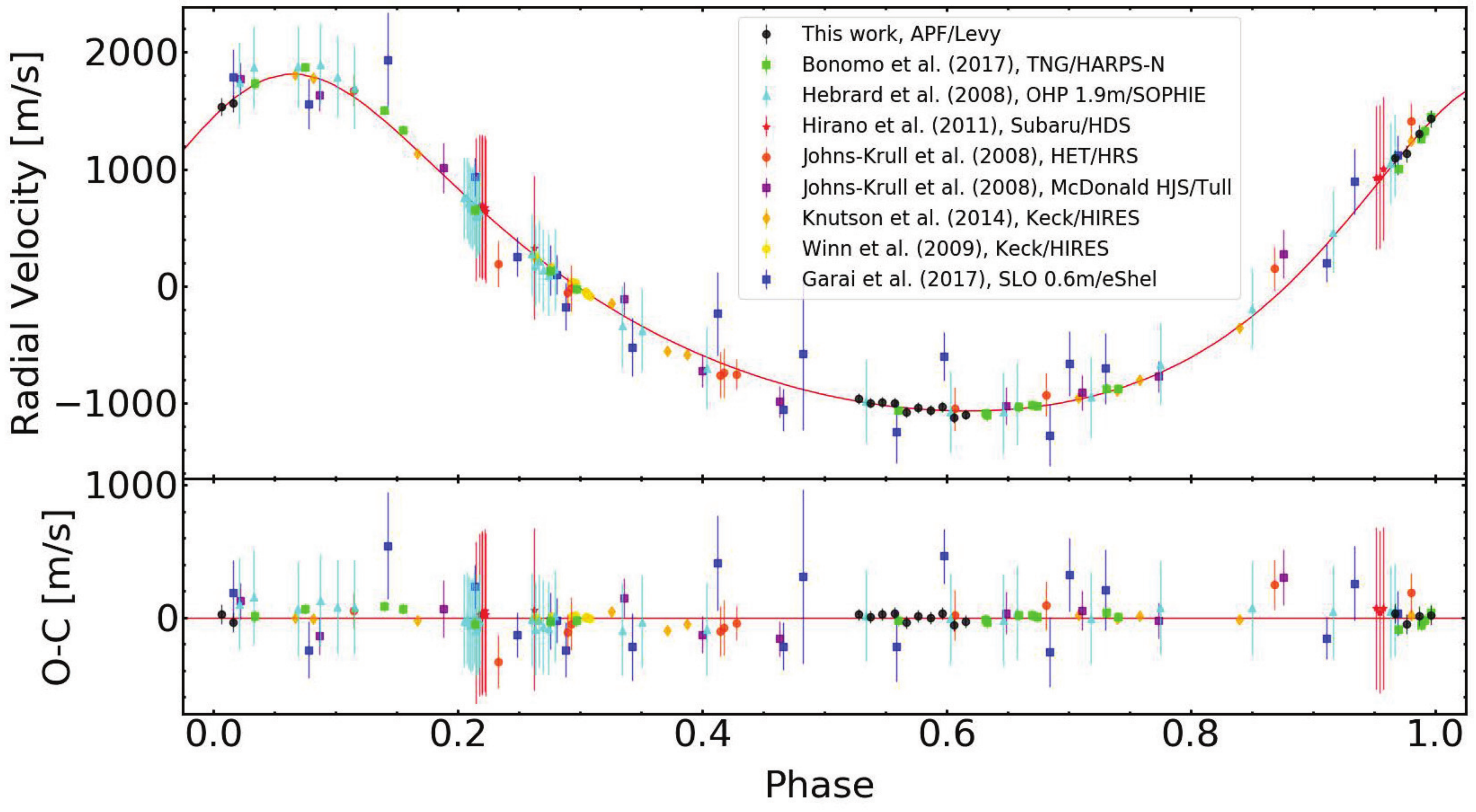}
\caption{The radial velocity measurements of XO-3 from previous works \citep{Hirano2011, Johns-Krull2008, Hebrard2008, Winn2009, Knutson2014, Bonomo2017, Garai2017} and from this work, marked as different colors and shown as a function of orbital phase. The best-fitting Keplerian orbit model (solid red line) is determined from the joint fitting of RVs and light curves. The residuals of the best fit, with an RMS scatter $\sigma=267.90\,\rm{m\,s^{-1}}$, is shown in the bottom panel. }
\label{f3}
\end{figure}

These APF velocities are combined with previously published RV data sets from the following instruments (Figure \ref{f3}): the HIgh Resolution Echelle Spectrometer (HIRES) on Keck I \citep{Vogt2014}; the northern High Accuracy Radial velocity Planet Searcher (HARPS-N) on the Telescopio Nazionale Galileo \citep{Cosentino2012, Cosentino2014}; the Spectrographe pour l'Observation des Ph$\mathrm{\acute{e}}$nom$\mathrm{\grave{e}}$nes des
Int$\mathrm{\acute{e}}$rieurs stellaires et des
Exoplan$\mathrm{\grave{e}}$tes (SOPHIE) on the 1.93m reflector telescope at the Haute-Provence Observatory \citep{Perruchot2008}; the High Dispersion Spectrograph (HDS) on the Subaru telescope \citep{Noguchi2002}; the High Resolution Spectrograph (HRS) on the Hobby-Eberly Telescope and the Tull Coude spectrograph on the HJS telescope (though, in this case, it was coupled to the Hobby-Eberly Telescope via a fiber optic cable) \citep{Tull1995, Tull1998}; and the eShel spectrograph at Stara Lesna Observatory (SLO) \citep{Eversberg2016}. Detailed information on each RV dataset can be found in the references list in Table \ref{RVsource_table}.

Like the APF and Keck/HIRES, HET's HRS makes use of an iodine cell for its wavelength calibration efforts and applies a forward modeling approach for determining each observation's RV measurement. The other instruments listed here handle wavelength calibration with reference spectra taken using a ThAr calibration lamp, either via simultaneous reference spectra in the case of fiber-fed instruments or via separate calibration frames for the slit-fed instruments. 

For the instruments with a ThAr-based wavelength calibration, a 2-D spectrum is extracted from the FITS file once an observation is complete. The stellar spectrum is cross-correlated with a reference (in the case of HARPS-N, SOPHIE, and HDS this is a numerical mask corresponding to the appropriate spectral type, while for the Tull Coude spectrograph it is a particular spectrum of XO-3 taken on BJD 2454137.8215). The resulting cross-correlation function (CCF) is fit with a Gaussian curve to produce a radial velocity measurement, and it is calibrated to determine the RV photon-noise uncertainty $\sigma_{RV}$ \citep[e.g.][]{Baranne1996, Pepe2002}.

\begin{deluxetable*}{lcccccc}
\tablecaption{RV data sources\label{RVsource_table}}
\tablehead{\colhead{~~~Reference} & \colhead{Facility} & \colhead{Calibration} & \colhead{ } & \colhead{  N$_{obs}$  }&  \colhead{}  & \colhead{RMS (m s$^{-1}$})}
\startdata
This work                     & Levy (APF)              & I$_{2}$ cell  & \hspace{.5cm} & 16  & \hspace{.5cm} & 29.95  \\
Bonomo et al. (2017)          & HARPS-N (TNG)           & ThAr lamp     &  & 19  &  & 44.13  \\
Hebrard et al. (2008)         & \hspace{.25cm} SOPHIE (1.93-m, HPO) \hspace{.25cm} & ThAr lamp     &  & 34  &  & 66.37  \\
Hirano et al. (2011)          & HDS (Subaru)            & ThAr lamp     &  & 10  &  & 47.64  \\
Johns-Krull et al. (2008) \hspace{.25cm}     & HRS (HET)               & I$_{2}$ cell  &  & 11  &  & 151.41 \\
Johns-Krull et al. (2008)     & Tull Coude (HJS)        & I$_{2}$ cell  &  & 10  &  & 141.73 \\
Knutson et al. (2014)         & HIRES (Keck I)          & I$_{2}$ cell  &  & 11  &  & 37.20  \\
Winn et al. (2009)            & HIRES  (Keck I)         & ThAr lamp     &  & 11  &  & 11.45  \\
Garai et al. (2017)           & eShel (0.6-m, SLO)      & ThAr lamp     &  & 20  &  & 269.30 \\
\enddata
\end{deluxetable*}

\section{Stellar Atmospheric Parameters from APF }
\label{sect3}

We determined the values of the stellar atmospheric parameters, including $V\sin i_{*}$, $T_{\rm eff}$, and \logg, for XO-3 directly from our APF spectrum using \textit{The Cannon} \citep{Ness2015, Casey2016}, a data-driven approach using generative modeling to determine stellar parameters. \textit{The Cannon} trains on an input set of stellar spectra with previously determined parameter values in order to ``learn" the characteristics of this dataset. This trained model can then be applied to obtain the corresponding labels for a new set of input spectra. \textit{The Cannon} requires overlapping wavelength coverage and assumes a similar set of systematics between the training set of spectra and the data to which it is applied.

Following the methods of \citet{Rice2020}, we ran \textit{The Cannon} to determine the properties of XO-3 directly from our APF spectrum. We trained our model using Keck/HIRES spectra included in the Spectral Properties of Cool Stars (SPOCS) dataset \citep{Valenti2005, Brewer2016}, with stellar parameters determined using the program Spectroscopy Made Easy (SME; \citealt{Valenti1996}).

From the full SPOCS sample, we removed all spectra flagged as `bad' or labeled with `NGC', indicating that the target was not an individual star. We also cut all spectra from our input sample with SNR $<$ 100, and we used only the highest-SNR spectrum for each star in the sample. This resulted in a total of 1202 spectra in our training/validation sample. We used 86 overlap stars observed by both Keck/HIRES (as part of the SPOCS sample) and by the APF as a validation set to test our model performance, leaving 1116 stars in our training sample.

We interpolated the HIRES and APF spectra onto an overlapping wavelength grid to directly compare each spectrum. Then, we trained our model using only HIRES spectra and applied it to our overlapping validation set of APF spectra to determine uncertainties in each parameter. Our final values for XO-3 are included in Table \ref{tab:stellar_params}.

Because we trained our model using HIRES spectra, the differing systematics between the Keck and APF instrumentation setups may affect the precision of our label transfer process. Indeed, we find that the scatter when transferring labels across datasets, quantified using our validation set, is higher than when using only Keck/HIRES spectra as in \citet{Rice2020}. This results in higher uncertainties for our reported values. From our validation set, we also find a systematic offset in the values of $V\sin i_{*}$ derived from our APF spectra relative to the nominal SPOCS values; that is, the $V\sin i_{*}$ values obtained for APF spectra with \textit{The Cannon} are on average -1.87 km/s lower than the associated SPOCS values. This offset is likely attributable to the difference in the line spread-function of the APF as compared to that of Keck/HIRES. We correct for this offset in Table \ref{tab:stellar_params}.

XO-3 is also included in the SPOCS sample with previously determined stellar parameters obtained based on the Keck spectra alone. We compared these pre-existing parameters with our results as an additional test to check the fidelity of our model, focusing on \logg\ and $T_{\rm eff}$, two other dominant global stellar properties that should show no temporal changes. As shown in Table \ref{tab:stellar_params}, we find that all global stellar parameters obtained for XO-3 from the APF data are in agreement with previously reported values derived from the HIRES dataset.

Our results are detailed in Table \ref{tab:stellar_params}. We report our uncertainties as the scatter in our test set results when testing and training with the SPOCS sample; these uncertainties may be underestimated due to the differing systematics across instruments, which are not encompassed by this scatter. Although our stellar atmospheric parameters from APF data show good agreement with previous estimates derived from a wide range of independent datasets, we adopt \logg\ and $T_{\rm eff}$ from \citet{Torres2012} as priors in our global fitting (see \S \ref{sect4} for details) because their parameters benefit from using combined high-SNR data from both Keck/HIRES and TRES/FIES.

Our primary parameter of interest is $V\sin i_{*}$ (see \S \ref{sect5} for details), which we compare with previous estimates in Table \ref{tab:stellar_params}. Ultimately, we find that the value of $V\sin i_{*}$ that we obtain directly from the APF spectrum is consistent with previous estimates derived from a wide range of independent datasets.

\section{Planetary Parameters from Global Fitting} \label{sect4}

\begin{deluxetable*}{lccccccc}
\tablewidth{0pt}
\tablecaption{\label{tab:stellar_params} Stellar parameters for XO-3}
\tablehead{
\colhead{Telescope}                      & \colhead{HJS}               & \colhead{Keck}         & \colhead{Keck}         & \colhead{Subaru}     & \colhead{Subaru+Keck}     & \colhead{Keck+TRES}      & \colhead{APF}      \\         
\colhead{Instrument}                     & \colhead{Tull Coude}        & \colhead{HIRES}        & \colhead{HIRES}        & \colhead{HDS}        & \colhead{HDS+HIRES}       & \colhead{HIRES+FIES}     & \colhead{Levy}      \\   
\colhead{Observation Date}               & 10/2006-02/2007             & 02/2009                & 02/2009                &  11/2009-02/2010     &  02/2009-02/2010          &  02/2009                 &      10/2016               \\
\colhead{Method}                         & \colhead{SME}               & \colhead{SME}          & \colhead{R-M Fitting}         & \colhead{R-M Fitting}      & \colhead{R-M Fitting}  & \colhead{SPC+SME+MOOG}              & \colhead{Cannon}     \\
\colhead{References}                     & \colhead{1}  & \colhead{2}   & \colhead{3}    & \colhead{4}   & \colhead{4}                & \colhead{5}        & \colhead{This work}   \\  
}
\startdata
$T_{\rm eff}$ (K)                        &  $6429  \pm 50$                      &  $6673   \pm 25$                  & ...                          & ...                            & ...                                         &    $6759 \pm 79$                      & $6430 \pm 69$   \\ 
$[M/H]$                                  & $-0.204 \pm 0.023$                   &  $-0.054 \pm 0.01$                & ...                          & ...                            & ...                                         &    $-0.05 \pm 0.08$                   & ...           \\
$\mathrm{log} g$ ($\rm{cm\,s^{-2}}$)               &  $3.95  \pm 0.062$                   &  $4.15   \pm 0.028$               & ...                          & ...                            & ...                                         &     $4.24 \pm 0.03^{\rm a}$                   & $4.3 \pm 0.1$ \\
$V\sin i_{*}$ ($\rm{km\,s^{-1}}$)    &  $18.54 \pm 0.17$                    &  $16.19  \pm 0.5$                 & $18.31 \pm 1.3$                & $17.0 \pm 1.2$                   & $18.4 \pm 0.8$                                &    $20.3 \pm 2.0$                     & $17.3 \pm 0.9$  \\
\enddata
\tablenotetext{}{$^{a}$External constraint on $\mathrm{log} g$ was obtained from the light curves \citep{Winn2008}.}
\tablenotetext{}{$^{1}$\citet{Johns-Krull2008}; $^{2}$\citet{Brewer2016}; $^{3}$\citet{Winn2009}; $^{4}$\citet{Hirano2011};  $^{5}$\citet{Torres2012}.}
\end{deluxetable*}

\textbf{System Parameters}. To determine the XO-3 system parameters, we used EXOFASTv2 \citep{eastman2013, eastman2017} to simultaneously fit the transit light curves and out-of-transit radial velocity data from the literature \citep{Johns-Krull2008, Winn2008, Hebrard2008, Winn2009, Bonomo2017, Hirano2011, Knutson2014, Bonomo2017, Garai2017}, as well as our newly collected photometric and RV data. 

EXOFASTv2 performs a global analysis of exoplanetary and stellar parameters using a differential evolution Markov Chain Monte Carlo (DE-MCMC, \citealt{Braak2006}) simulation to simultaneously fit, for an arbitrary number of planets, the Spectral Energy Distribution (SED), transit data, and RV data taken from multiple instruments.

To constrain the stellar parameters, we used the MESA Isochrones and Stellar Tracks (MIST) model \citep{Choi2016, Dotter2016} included in EXOFASTv2. The Gaussian priors were applied to the $T_{\rm eff}$ and \feh\ of the star derived in \citet{Torres2012}. The limb-darkening coefficients were assumed to be a quadratic function. We imposed wavelength-dependent priors on limb-darkening coefficients from \citet{Claret2011} based on the \teff, \logg, and \feh\ from \citet{Torres2012}. The priors for the orbital parameters, including all transit and RV parameters, were adopted from the results of \citet{Winn2009}.

To minimize the convergence time, the fitting process required multiple short runs before longer ones. We derived a new set of Gaussian priors after each run, allowing us to begin subsequent fits at the most likely model. We continually refined the fit until our criteria -- both the number of independent draws being greater than 1000 and a Gelman-Rubin statistic of less than 1.01 for all parameters -- were satisfied six consecutive times, indicating that the chains were considered to be well-mixed \citep{eastman2013}. 

The system parameters derived from global fit are listed in Table ~\ref{tab:XO3global}. The fitting results are shown in Figure~\ref{f1}, \ref{f2} and \ref{f3}.

\begin{figure}[t]
\centering
\includegraphics[width=1\columnwidth]{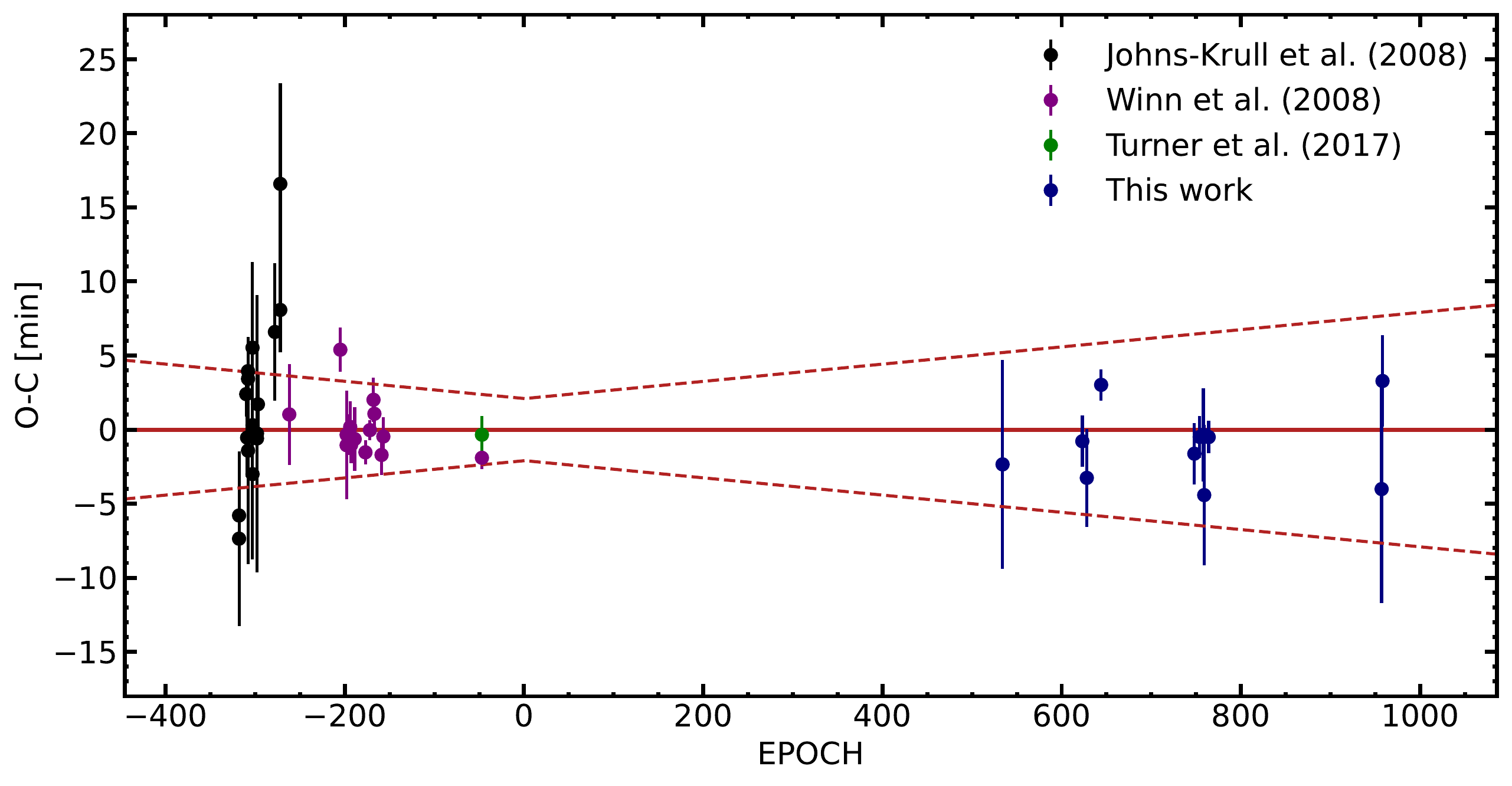}
\caption{Observed minus calculated mid-transit times (O-C) for XO-3b, according to the updated linear ephemeris. The $(O-C)=0$ reference is represented by a red solid line, and the red dashed line indicates the propagation of $\pm$ 1 $\sigma$ errors of the updated linear ephemeris.}
\label{f4}
\end{figure}

\textbf{Transit Timing Variations $\&$ Orbital Ephemeris}. We modelled each available transit light curve for XO-3 using the JKTEBOP \citep{Southworth2008} code. We fixed all global parameters to the results derived from the global fitting, and we allowed only the transit mid-time ($T_0$) and baseline flux ($F_{0}$) to vary as free parameters in the fit. We utilized the bootstrapping technique, Monte Carlo simulations, and the residual-shift method to estimate the errors of mid-transit times separately. The largest errors were selected as the final errors to provide a conservative estimate. The result is shown in Figure~\ref{f4}, which is consistent with a constant period. No significant transit timing variations were detected.

To update the linear ephemeris ($T_{\rm C}+N \times P$), we performed a weighted least squares fit to the derived mid-transit times ($T_0$). During the fit, we followed the approaches described in \cite{Southworth2017}, and we re-scaled the uncertainties of each transit mid-time such that $\chi^{2}_{\rm reduced}=1$. This choice was made to provide conservative errors for the transit mid-time at the reference epoch ($T_{\rm C}$) and orbital period ($P$) for future scheduling purposes. The result agrees with the values from \citet{Winn2008} within 1$\sigma$.

\startlongtable
\begin{deluxetable*}{lccccc}
\tablecaption{System Parameters for XO-3}
\tablehead{\colhead{~~~Parameter}            & \colhead{Units}                             & \colhead{This work} & \colhead{Previous work} & \colhead{Agreement($\sigma$)}& \colhead{Ref}}
\startdata
\sidehead{Stellar Parameters:}                
~~~~$M_*$\dotfill                          &Mass (\msun)\dotfill                              &$1.219^{+0.090}_{-0.095}$          &$1.213\pm 0.066$                 &0.05        &Winn2008             \\                                   
~~~~$R_*$\dotfill                          &Radius (\rsun)\dotfill                            &$1.371^{+0.041}_{-0.042}$          &$1.377 \pm 0.083$                &0.06        &Winn2008             \\                                   
~~~~$L_*$\dotfill                          &Luminosity (\lsun)\dotfill                        &$2.97^{+0.27}_{-0.26}$             &$2.92^{+0.59}_{-0.48}$           &0.09        &Winn2008             \\                                   
~~~~$\rho_*$\dotfill                       &Density (cgs)\dotfill                             &$0.665^{+0.033}_{-0.030}$          &$0.650\pm 0.086$                 &0.16        &Winn2008             \\                                   
~~~~$\log{g}$\dotfill                      &Surface gravity (cgs)\dotfill                     &$4.249^{+0.017}_{-0.018}$          &$4.244\pm 0.041$                 &0.11        &Winn2008             \\                                   
~~~~$T_{\rm eff}$\dotfill                  &Effective Temperature (K)\dotfill                 &$6471^{+83}_{-82}$                 &$6429\pm 100$                    &0.32        &Winn2008             \\                                   
~~~~$[{\rm Fe/H}]$\dotfill                 &Metallicity \dotfill                               &$-0.176^{+0.080}_{-0.081}$         &$-0.177\pm 0.080$                &0.01        &Winn2008             \\                                   
~~~~$[{\rm Fe/H}]_{0}$\dotfill             &Initial Metallicity \dotfill                       &$-0.025^{+0.072}_{-0.071}$         &...                              &...         &...                  \\                                   
~~~~$Age$\dotfill                          &Age (Gyr)\dotfill                                 &$2.6^{+1.6}_{-1.1}$                &$2.82^{+0.58}_{-0.82}$           &0.18        &Winn2008             \\                                   
~~~~$EEP$\dotfill                          &Equal Evolutionary Point \dotfill                 &$361^{+39}_{-21}$                  &...                              &...         &...                  \\                                   
~~~~$A_v$\dotfill                          &V-band extinction \dotfill                        &$0.069^{+0.091}_{-0.049}$          &...                              &...         &...                  \\                                   
~~~~$\sigma_{\rm SED}$\dotfill                 &SED photometry error scaling \dotfill             &$2.8^{+2.8}_{-1.1}$                &...                              &...         &...                  \\                                   
~~~~$d$\dotfill                            &Distance (pc)\dotfill                             &$178.6^{+6.9}_{-7.4}$              &$174  \pm 18$                    &0.24        &Winn2008             \\                                   
~~~~$\pi$\dotfill                          &Parallax (mas)\dotfill                            &$5.60^{+0.24}_{-0.21}$             &...                              &...         &...                  \\                                   
\sidehead{Planetary Parameters:}                                                                                                                                                                          
~~~~$P$\dotfill                           &Period (days)\dotfill                              &$3.19152\pm0.00145^{a}$          &$3.1915239\pm 0.0000068$             &0.00    &Winn2008             \\                                 
~~~~$R_P$\dotfill                         &Radius (\rj)\dotfill                               &$1.219^{+0.039}_{-0.040}$          &...                                  &...     &...                  \\                                 
~~~~$T_C$\dotfill                         &Time of Transit (\bjdtdb)\dotfill                  &$2454449.86969\pm0.00073^{a}$          &$2454449.868937 \pm 0.00023$          &0.99    &Winn2008             \\                                 
~~~~$T_0$\dotfill                         &Optimal Transit Time (\bjdtdb)\dotfill             &$2455314.77290\pm0.00015$          &...                                  &...     &...                  \\                                 
~~~~$a$\dotfill                           &Semi-major axis (AU)\dotfill                       &$0.0455^{+0.0011}_{-0.0012}$       &$0.04539^{+0.00081}_{-0.00084}$      &0.08    &Bonomo2017           \\                                 
~~~~$i$\dotfill                           &Inclination (Degrees)\dotfill                      &$84.26\pm0.19$                     &$84.20  \pm 0.54$                    &0.1     &win2008              \\                                 
~~~~$e$\dotfill                           &Eccentricity \dotfill                              &$0.2853^{+0.0027}_{-0.0026}$       &$0.27587^{+0.00071}_{-0.00067}$      &3.39    &Bonomo2017           \\                                 
~~~~$\omega_*$\dotfill                    &Argument of Periastron (Degrees)\dotfill           &$349.6\pm1.2$                      &$349.35^{+0.67}_{-0.68}$             &1.78  &Bonomo2017           \\                                 
~~~~$T_{\mathrm{eq}}$\dotfill                      &Equilibrium temperature (K)\dotfill                &$1714\pm26$                        &$1710 \pm 46$                        &0.08    &Winn2008             \\                                 
~~~~$M_P$\dotfill                         &Mass (\mj)\dotfill                                 &$11.92^{+0.59}_{-0.63}$            &$11.7^{+0.43}_{-0.43}$               &0.3     &Bonomo2017           \\                                 
~~~~$K$\dotfill                           &RV semi-amplitude (m/s)\dotfill                    &$1488.0^{+8.7}_{-9.1}$             &$1468.9^{+4.6}_{-4.5}$               &1.95    &Bonomo2017           \\                                 
~~~~$\mathrm{log}K$\dotfill                        &Log of RV semi-amplitude \dotfill                  &$3.1726^{+0.0025}_{-0.0026}$       & ...                                 &...     &...                  \\                                 
~~~~$R_P/R_*$\dotfill                     &Radius of planet in stellar radii \dotfill         &$0.09139^{+0.00039}_{-0.00040}$    &$0.09057   \pm 0.00057$              &1.19    &Winn2008             \\                                     
~~~~$a/R_*$\dotfill                        &Semi-major axis in stellar radii \dotfill         &$7.12^{+0.12}_{-0.11}$             &$7.07   \pm 0.31$                    &0.15    &Winn2008             \\                                     
~~~~$\delta$\dotfill                       &Transit depth (fraction)\dotfill                  &$0.008352^{+0.000071}_{-0.000072}$ &...                                  &...     &...                  \\                                     
~~~~$$Depth$$\dotfill                        &Flux decrement at mid transit \dotfill            &$0.008352^{+0.000071}_{-0.000072}$ &...                                  &...     &...                  \\                                     
~~~~$\tau$\dotfill                         &Ingress/egress transit duration (days)\dotfill    &$0.01904\pm0.00067$                &$0.466 \pm 0.033$                    &0.25   & Winn2008            \\                                     
~~~~$T_{14}$\dotfill                       &Total transit duration (days)\dotfill             &$0.12334\pm0.00060$                &...                                  &...     & ...                 \\                                     
~~~~$T_{\mathrm{FWHM}}$\dotfill                     &FWHM transit duration (days)\dotfill              &$0.10430\pm0.00033$                &...                                  &...     & ...                 \\                                     
~~~~$b$\dotfill                            &Transit Impact parameter \dotfill                 &$0.700^{+0.011}_{-0.012}$          &$0.705 \pm 0.023$                    &0.19    & Winn2008            \\                                     
~~~~$b_S$\dotfill                          &Eclipse impact parameter \dotfill                 &$0.614^{+0.012}_{-0.013}$          &...                                  &...     &...                  \\                                     
~~~~$\tau_S$\dotfill                       &Ingress/egress eclipse duration (days)\dotfill    &$0.01509^{+0.00053}_{-0.00052}$    &...                                  &...     &...                  \\                                     
~~~~$T_{S,14}$\dotfill                     &Total eclipse duration (days)\dotfill             &$0.1167\pm0.0012$                  &...                                  &...     &...                  \\                                     
~~~~$T_{S,\mathrm{FWHM}}$\dotfill                   &FWHM eclipse duration (days)\dotfill              &$0.10159^{+0.00069}_{-0.00075}$    &...                                  &...     &...                  \\                                     
~~~~$\delta_{S,3.6\mu m}$\dotfill          &Blackbody eclipse depth at 3.6$\mu$m (ppm)\dotfill&$754^{+24}_{-25}$                  &...                                  &...     &...                  \\                                     
~~~~$\delta_{S,4.5\mu m}$\dotfill          &Blackbody eclipse depth at 4.5$\mu$m (ppm)\dotfill&$975\pm27$                         &...                                  &...     &...                  \\                                     
~~~~$\rho_P$\dotfill                       &Density (cgs)\dotfill                             &$8.15^{+0.54}_{-0.50}$             &$8.1^{+1.6}_{-1.3}$                  &0.04    &Bonomo2017           \\                                     
~~~~$\mathrm{log}g_P$\dotfill                       &Surface gravity \dotfill                          &$4.298^{+0.017}_{-0.016}$          &$4.293^{+0.055}_{-0.053}$            &0.09    &Bonomo2017           \\                                     
~~~~$\Theta$\dotfill                       &Safronov Number \dotfill                          &$0.729^{+0.025}_{-0.023}$          &...                                  &...     &...             \\                                     
~~~~$\fave$\dotfill                        &Incident Flux (\fluxcgs)\dotfill                  &$1.81\pm0.11$                      &...                                  &...     &...                  \\                                     
~~~~$T_P$\dotfill                          &Time of Periastron (\bjdtdb)\dotfill              &$2454449.2483^{+0.0086}_{-0.0089}$ &...                                  &...     &...                  \\                                     
~~~~$T_S$\dotfill                          &Time of eclipse (\bjdtdb)\dotfill                 &$2454448.8324^{+0.0036}_{-0.0038}$ &...                                  &...     &...                  \\                                     
~~~~$T_A$\dotfill                          &Time of Ascending Node (\bjdtdb)\dotfill          &$2454449.3105^{+0.0035}_{-0.0037}$ &...                                  &...     &...                  \\                                     
~~~~$T_D$\dotfill                          &Time of Descending Node (\bjdtdb)\dotfill         &$2454447.852\pm0.010$              &...                                  &...     &...                  \\                                     
~~~~$e\cos{\omega_*}$\dotfill               & \dotfill                                         &$0.2778^{+0.0018}_{-0.0019}$       &$0.27111_{-0.00033}^{+0.00034}$      &3.65    &Bonomo2017           \\                                     
~~~~$e\sin{\omega_*}$\dotfill               & \dotfill                                         &$-0.0648^{+0.0061}_{-0.0063}$      &$-0.0510_{-0.0034}^{+0.0033}$        &1.93    & Bonomo2017          \\                                     
~~~~$M_P\sin i$\dotfill                    &Minimum mass (\mj)\dotfill                        &$11.86^{+0.58}_{-0.63}$            &...                                  &...     &...                  \\                                     
~~~~$M_P/M_*$\dotfill                      &Mass ratio \dotfill                               &$0.00934^{+0.00026}_{-0.00023}$    &$0.00927 \pm 0.00036$                &0.16    &Winn2008             \\                                     
~~~~$d/R_*$\dotfill                        &Separation at mid transit \dotfill                &$7.00\pm0.13$                      &...                                  &...     &...                  \\                                     
~~~~$P_T$\dotfill                          &A priori non-grazing transit prob \dotfill        &$0.1298^{+0.0023}_{-0.0024}$       &...                                  &...     &...                  \\                                     
~~~~$P_{T,G}$\dotfill                      &A priori transit prob \dotfill                    &$0.1559\pm0.0029$                  &...                                  &...     &...                  \\                                     
~~~~$P_S$\dotfill                          &A priori non-grazing eclipse prob \dotfill        &$0.1478^{+0.0021}_{-0.0022}$       &...                                  &...     &...                  \\                                     
~~~~$P_{S,G}$\dotfill                      &A priori eclipse prob \dotfill                    &$0.1776\pm0.0027$                  &...                                  &...     &...                  \\                                     
\sidehead{Wavelength Parameters:} 
~~~~$u_{1B}$\dotfill                      &linear limb-darkening coeff \dotfill               &$0.497\pm0.028$                    & ...                                 & ...    &...                  \\                                          
~~~~$u_{2B}$\dotfill                      &quadratic limb-darkening coeff \dotfill            &$0.277\pm0.026$                    & ...                                 & ...    &...                  \\                                                 
~~~~$u_{1I}$\dotfill                      &linear limb-darkening coeff \dotfill               &$0.197\pm0.017$                    &$0.06\pm0.15$                                 & 0.91   &Winn2008             \\                                          
~~~~$u_{2I}$\dotfill                      &quadratic limb-darkening coeff \dotfill            &$0.301\pm0.016$                    &0.35$^{b}$                                 & ...   &Winn2008             \\                                              
~~~~$u_{1R}$\dotfill                      &linear limb-darkening coeff \dotfill               &$0.260^{+0.013}_{-0.014}$          &$0.16\pm0.14$                                 & 0.71   &Winn2008             \\                                          
~~~~$u_{2R}$\dotfill                      &quadratic limb-darkening coeff \dotfill            &$0.317\pm0.012$                    &0.37$^{b}$                                 & ...   &Winn2008             \\                                                                                                                                         
~~~~$u_{1\mathrm{Sloan}r}$\dotfill                 &linear limb-darkening coeff \dotfill               &$0.293\pm0.044$                    &...                                  & ...    &...                  \\                                          
~~~~$u_{2\mathrm{Sloan}r}$\dotfill                 &quadratic limb-darkening coeff \dotfill            &$0.336^{+0.046}_{-0.045}$          &...                                  & ...    &...                  \\                                                
~~~~$u_{1\mathrm{Sloan}z}$\dotfill                 &linear limb-darkening coeff \dotfill               &$0.174\pm0.019$                    &$0.11\pm0.07$                                 & 0.88   &Winn2008             \\                                          
~~~~$u_{2\mathrm{Sloan}z}$\dotfill                 &quadratic limb-darkening coeff \dotfill            &$0.301\pm0.019$                    &0.353$^{b}$                                & ...   &Winn2008             \\                                                                                                                                            
~~~~$u_{1V}$\dotfill                      &linear limb-darkening coeff \dotfill               &$0.366^{+0.020}_{-0.019}$                    &$0.47\pm0.14$                                & 0.74   &Winn2008             \\                                         
~~~~$u_{2V}$\dotfill                      &quadratic limb-darkening coeff \dotfill            &$0.327\pm0.019$                    &0.36$^{b}$                                 & ...   &Winn2008             \\
\enddata
\label{tab:XO3global}
\tablenotetext{}{$^{a}$ Conservative values from JKTEBOP fit.}
\tablenotetext{}{$^{b}$ In the fit of \cite{Winn2008}, the quadratic coefficient is fixed.}
\end{deluxetable*}

\startlongtable
\begin{deluxetable*}{lcccc}
\tablecaption{Comparison of the System Parameters before \citet{Hebrard2008} and after \citet{Winn2009}}
\tablehead{\colhead{~~~Parameter} & \colhead{Units} &\colhead{Before \citet{Hebrard2008}}  & \colhead{After \citet{Winn2009}}& \colhead{Agreement($\sigma$)}}
\startdata
\sidehead{Stellar Parameters:} 
~~~~$M_*$\dotfill &Mass (\msun)\dotfill &$1.225^{+0.090}_{-0.097}$                                                                     &$1.216^{+0.087}_{-0.092}$                       &0.07                \\
~~~~$R_*$\dotfill &Radius (\rsun)\dotfill &$1.395\pm0.060$                                                                             &$1.325\pm0.050$                                 &0.9                 \\
~~~~$L_*$\dotfill &Luminosity (\lsun)\dotfill &$3.07^{+0.34}_{-0.32}$                                                                  &$2.81^{+0.29}_{-0.27}$                          &0.6                 \\
~~~~$\rho_*$\dotfill &Density (cgs)\dotfill &$0.634^{+0.067}_{-0.059}$                                                                 &$0.734^{+0.061}_{-0.055}$                       &1.18                \\
~~~~$\log{g}$\dotfill &Surface gravity (cgs)\dotfill &$4.236\pm0.030$                                                                  &$4.277\pm0.024$                                 &1.07                \\
~~~~$T_{\rm eff}$\dotfill &Effective Temperature (K)\dotfill &$6467^{+83}_{-81}$                                                       &$6490^{+83}_{-79}$                              &0.2                 \\
~~~~$[{\rm Fe/H}]$\dotfill &Metallicity \dotfill &$-0.174^{+0.079}_{-0.078}$                                                            &$-0.179^{+0.081}_{-0.082}$                      &0.04                \\
~~~~$[{\rm Fe/H}]_{0}$\dotfill &Initial Metallicity \dotfill &$-0.021^{+0.072}_{-0.071}$                                                &$-0.041\pm0.070$                                &0.2                 \\
~~~~$Age$\dotfill &Age (Gyr)\dotfill &$2.7^{+1.6}_{-1.1}$                                                                              &$2.2^{+1.6}_{-1.0}$                             &0.26                \\
~~~~$EEP$\dotfill &Equal Evolutionary Point \dotfill &$365^{+38}_{-23}$                                                                &$349^{+38}_{-18}$                               &0.38                \\
~~~~$A_v$\dotfill &V-band extinction \dotfill &$0.068^{+0.092}_{-0.048}$                                                               &$0.072^{+0.090}_{-0.051}$                       &0.04                \\
~~~~$\sigma_{\mathrm{SED}}$\dotfill &SED photometry error scaling \dotfill &$2.8^{+2.8}_{-1.1}$                                                 &$2.6^{+2.5}_{-1.0}$                             &0.07                \\
~~~~$d$\dotfill &Distance (pc)\dotfill &$181.6^{+8.8}_{-9.0}$                                                                          &$173.4^{+7.6}_{-7.8}$                           &0.7                 \\
~~~~$\pi$\dotfill &Parallax (mas)\dotfill &$5.51^{+0.29}_{-0.25}$                                                                      &$5.77^{+0.27}_{-0.24}$                          &0.71                \\
\sidehead{Planetary Parameters: }                                                                                                                                                                     
~~~~$P$\dotfill &Period (days)\dotfill &$3.1915219\pm0.0000035$                                                                        &$3.1915264^{+0.0000020}_{-0.0000019}$           &1.12                         \\
~~~~$R_P$\dotfill &Radius (\rj)\dotfill &$1.229^{+0.055}_{-0.056}$                                                                     &$1.183^{+0.050}_{-0.049}$                       &0.62                         \\
~~~~$T_C$\dotfill &Time of Transit (\bjdtdb)\dotfill &$2454449.86996^{+0.00025}_{-0.00026}$                                            &$2454449.8685\pm0.0016$                         &0.9                          \\
~~~~$T_0$\dotfill &Optimal Transit Time (\bjdtdb)\dotfill &$2454357.31583\pm0.00023$                                                   &$2454858.3839\pm0.0016$                         &.                    \\
~~~~$a$\dotfill &Semi-major axis (AU)\dotfill &$0.0455^{+0.0011}_{-0.0012}$                                                            &$0.0454^{+0.0011}_{-0.0012}$                    &0.06                         \\
~~~~$i$\dotfill &Inclination (Degrees)\dotfill &$84.08^{+0.38}_{-0.40}$                                                                &$84.67^{+0.31}_{-0.30}$                         &1.17                         \\
~~~~$e$\dotfill &Eccentricity \dotfill &$0.261\pm0.015$                                                                                &$0.2887^{+0.0061}_{-0.0059}$                    &1.71                         \\
~~~~$\omega_*$\dotfill &Argument of Periastron (Degrees)\dotfill &$-11.9^{+6.3}_{-6.0}$                                                &$-15.4\pm2.3$                                   &0.52                         \\
~~~~$T_{\rm eq}$\dotfill &Equilibrium temperature (K)\dotfill &$1727\pm35$                                                                 &$1691\pm31$                                     &0.77                         \\
~~~~$M_P$\dotfill &Mass (\mj)\dotfill &$11.93^{+0.69}_{-0.71}$                                                                         &$11.56^{+0.59}_{-0.63}$                         &0.40                          \\
~~~~$K$\dotfill &RV semi-amplitude (m/s)\dotfill &$1474\pm45$                                                                          &$1449^{+26}_{-27}$                              &0.48                         \\
~~~~$\mathrm{log}K$\dotfill &Log of RV semi-amplitude \dotfill &$3.169\pm0.013$                                                                 &$3.1612^{+0.0077}_{-0.0082}$                    &0.51                         \\
~~~~$R_P/R_*$\dotfill &Radius of planet in stellar radii \dotfill &$0.09050^{+0.00052}_{-0.00054}$                                     &$0.09174^{+0.00063}_{-0.00064}$                 &1.49                         \\
~~~~$a/R_*$\dotfill &Semi-major axis in stellar radii \dotfill &$7.01^{+0.24}_{-0.23}$                                                 &$7.36^{+0.20}_{-0.19}$                          &1.15                         \\
~~~~$\delta$\dotfill &Transit depth (fraction)\dotfill &$0.008191^{+0.000095}_{-0.000097}$                                             &$0.00842\pm0.00012$                             &1.48                         \\
~~~~$Depth$\dotfill &Flux decrement at mid transit \dotfill &$0.008191^{+0.000095}_{-0.000097}$                                        &$0.00842\pm0.00012$                             &1.48                         \\
~~~~$\tau$\dotfill &Ingress/egress transit duration (days)\dotfill &$0.01948^{+0.00089}_{-0.00090}$                                    &$0.0182\pm0.0010$                               &0.96                         \\
~~~~$T_{14}$\dotfill &Total transit duration (days)\dotfill &$0.12298^{+0.00083}_{-0.00085}$                                           &$0.12315^{+0.00094}_{-0.00093}$                 &0.13                         \\
~~~~$T_{\rm FWHM}$\dotfill &FWHM transit duration (days)\dotfill &$0.10350\pm0.00046$                                                      &$0.10495^{+0.00053}_{-0.00054}$                 &2.07                         \\
~~~~$b$\dotfill &Transit Impact parameter \dotfill &$0.713^{+0.013}_{-0.015}$                                                          &$0.679^{+0.019}_{-0.021}$                       &1.38                         \\
~~~~$b_S$\dotfill &Eclipse impact parameter \dotfill &$0.640^{+0.039}_{-0.037}$                                                        &$0.582^{+0.022}_{-0.023}$                       &1.28                         \\
~~~~$\tau_S$\dotfill &Ingress/egress eclipse duration (days)\dotfill &$0.0159^{+0.0018}_{-0.0015}$                                     &$0.01404^{+0.00084}_{-0.00080}$                 &0.94                         \\
~~~~$T_{S,14}$\dotfill &Total eclipse duration (days)\dotfill &$0.1182^{+0.0031}_{-0.0036}$                                            &$0.1142^{+0.0022}_{-0.0023}$                    &1.04                         \\
~~~~$T_{S,\mathrm{FWHM}}$\dotfill &FWHM eclipse duration (days)\dotfill &$0.1023^{+0.0012}_{-0.0021}$                                           &$0.1001^{+0.0014}_{-0.0016}$                    &1.1                          \\
~~~~$\delta_{S,3.6\mu m}$\dotfill &Blackbody eclipse depth at 3.6$\mu$m (ppm)\dotfill &$754\pm38$                                      &$731^{+36}_{-35}$                               &0.45                         \\
~~~~$\delta_{S,4.5\mu m}$\dotfill &Blackbody eclipse depth at 4.5$\mu$m (ppm)\dotfill &$973\pm42$                                      &$951\pm40$                                      &0.38                         \\
~~~~$\rho_P$\dotfill &Density (cgs)\dotfill &$7.97^{+1.0}_{-0.90}$                                                                     &$8.64^{+0.93}_{-0.83}$                          &0.52                         \\
~~~~$\mathrm{log}g_P$\dotfill &Surface gravity \dotfill &$4.292^{+0.035}_{-0.036}$                                                              &$4.310^{+0.029}_{-0.028}$                       &0.39                         \\
~~~~$\Theta$\dotfill &Safronov Number \dotfill &$0.722^{+0.043}_{-0.041}$                                                              &$0.730^{+0.034}_{-0.033}$                       &0.15                         \\
~~~~$\fave$\dotfill &Incident Flux (\fluxcgs)\dotfill &$1.89^{+0.16}_{-0.15}$                                                          &$1.71^{+0.13}_{-0.12}$                          &0.9                          \\
~~~~$T_P$\dotfill &Time of Periastron (\bjdtdb)\dotfill &$2454449.234^{+0.050}_{-0.051}$                                               &$2454449.230\pm0.016$                           &0.08                         \\
~~~~$T_S$\dotfill &Time of eclipse (\bjdtdb)\dotfill &$2454448.785^{+0.026}_{-0.028}$                                                  &$2454448.8326^{+0.0074}_{-0.0078}$              &1.64                         \\
~~~~$T_A$\dotfill &Time of Ascending Node (\bjdtdb)\dotfill &$2454449.293^{+0.021}_{-0.023}$                                           &$2454449.3025^{+0.0066}_{-0.0067}$              &0.40                          \\
~~~~$T_D$\dotfill &Time of Descending Node (\bjdtdb)\dotfill &$2454447.809^{+0.045}_{-0.046}$                                          &$2454447.869^{+0.021}_{-0.020}$                 &1.19                         \\
~~~~$ecos{\omega_*}$\dotfill & \dotfill &$0.254^{+0.013}_{-0.014}$                                                                     &$0.2784^{+0.0038}_{-0.0041}$                    &1.68                         \\
~~~~$esin{\omega_*}$\dotfill & \dotfill &$-0.053\pm0.029$                                                                              &$-0.077\pm0.012$                                &0.76                         \\
~~~~$M_P\sin i$\dotfill &Minimum mass (\mj)\dotfill &$11.87^{+0.68}_{-0.71}$                                                           &$11.51^{+0.59}_{-0.62}$                         &0.39                         \\
~~~~$M_P/M_*$\dotfill &Mass ratio \dotfill &$0.00932^{+0.00038}_{-0.00036}$                                                            &$0.00909^{+0.00029}_{-0.00027}$                 &0.49                         \\
~~~~$d/R_*$\dotfill &Separation at mid transit \dotfill &$6.90^{+0.40}_{-0.38}$                                                        &$7.31^{+0.24}_{-0.23}$                          &0.91                         \\
~~~~$P_T$\dotfill &A priori non-grazing transit prob \dotfill &$0.1317^{+0.0077}_{-0.0073}$                                            &$0.1243\pm0.0039$                               &0.86                         \\
~~~~$P_{T,G}$\dotfill &A priori transit prob \dotfill &$0.1579^{+0.0093}_{-0.0088}$                                                    &$0.1494^{+0.0049}_{-0.0048}$                    &0.81                         \\
~~~~$P_S$\dotfill &A priori non-grazing eclipse prob \dotfill &$0.1466^{+0.0033}_{-0.0034}$                                            &$0.1450^{+0.0034}_{-0.0035}$                    &0.33                         \\
~~~~$P_{S,G}$\dotfill &A priori eclipse prob \dotfill &$0.1758\pm0.0041$                                                               &$0.1743^{+0.0043}_{-0.0044}$                    &0.25                         \\
\sidehead{Wavelength Parameters:}                                                                                                                                                                    
 ~~~~$u_{1B}$\dotfill        &linear limb-darkening coeff \dotfill        &$0.500\pm0.032$                                             &$0.488\pm0.050$                                 &0.2                    \\
 ~~~~$u_{2B}$\dotfill        &quadratic limb-darkening coeff \dotfill     &$0.277\pm0.030$                                             &$0.277\pm0.050$                                 &0.0                    \\
 ~~~~$u_{1I}$\dotfill        &linear limb-darkening coeff \dotfill        &$0.190\pm0.018$                                             &$0.221\pm0.042$                                 &0.68                   \\
 ~~~~$u_{2I}$\dotfill        &quadratic limb-darkening coeff \dotfill     &$0.299^{+0.017}_{-0.018}$                                   &$0.300\pm0.044$                                 &0.02                   \\                                                                     
 ~~~~$u_{1R}$\dotfill        &linear limb-darkening coeff \dotfill        &$0.272^{+0.019}_{-0.020}$                                   &$0.258\pm0.016$                                 &0.56                   \\                                                                          
 ~~~~$u_{2R}$\dotfill        &quadratic limb-darkening coeff \dotfill     &$0.324\pm0.018$                                             &$0.319\pm0.015$                                 &0.21                   \\                                                                    
 ~~~~$u_{1\mathrm{Sloan}z}$\dotfill   &linear limb-darkening coeff \dotfill        &$0.166\pm0.021$                                             &...                                 &...                    \\                                                                    
 ~~~~$u_{2\mathrm{Sloan}z}$\dotfill   &quadratic limb-darkening coeff \dotfill     &$0.294\pm0.021$                                             &...                                  &...                   \\       
  ~~~~$u_{1\mathrm{Sloan}i}$\dotfill   &linear limb-darkening coeff \dotfill        &...                                            &$0.306\pm0.044$                                 &...                 \\                                                                    
 ~~~~$u_{2\mathrm{Sloan}i}$\dotfill   &quadratic limb-darkening coeff \dotfill     &...                                         &$0.340\pm0.045$                                 &...                    \\                                                                  
 ~~~~$u_{1V}$\dotfill        &linear limb-darkening coeff \dotfill        &$0.364\pm0.023$                                             &$0.341^{+0.049}_{-0.050}$                       &0.42                   \\                                                                    
 ~~~~$u_{2V}$\dotfill        &quadratic limb-darkening coeff \dotfill     &$0.324^{+0.022}_{-0.021}$                                   &$0.310\pm0.049$                                 &0.26                   \\                        
\enddata                        
\label{tab:XO3beforeafter}       
\end{deluxetable*}

\section{Discussion} \label{sect5}

Our global analysis of 45 transit light curves (including 12 collected in this work)  and 142 Doppler velocities (including 16 collected in this work) spans more than 10 years, making XO-3b one of the best-studied exoplanets.

The results from our global analysis (Table~\ref{tab:XO3global}) show good agreement with previous work \citep{Winn2009, Wong2014, Bonomo2017}, except that we find a slightly higher eccentricity ($e$) by $0.01$.

Our result confirms that the XO-3 system is unique, containing a massive planet ($M_{\rm P}=11.92^{+0.59}_{-0.63}\,{\rm \mj}$) in a relatively eccentric ($e=0.2853^{+0.0027}_{-0.0026}$) and short-period ($3.19152 \pm 0.00145\,$day) orbit around a massive star ($M_*=1.219^{+0.090}_{-0.095} M_{\odot}$). 

Our result also confirms the relatively high Safronov number, $\Theta=0.73$, of XO-3b \citep{Safronov1972}. The Safronov number is defined as
\begin{equation}
\Theta={{1}\over{2}}[ {{V_{\rm esc}}\over{V_{\rm orb}}} ]^2={M_{\rm P}\over{M_{*}}} {{a}\over{R_{\rm P}}},
\label{eq:K}
\end{equation}
where $V_{\rm esc}=\sqrt{2GM_{\rm P}/R_{\rm P}}$ is the planetary escape velocity, and $V_{\rm orb}=\sqrt{GM_{*}/a}$ is the planet's circular orbital velocity. $\Theta$ connects to the outcome of instability in an N-body system (for example, a system that hosts a hot Jupiter and close-in test particles). If $\Theta>>1$, the ejection of test particles is very likely when instability occurs, whereas for most hot Jupiters, $\Theta$ is substantially smaller than 1, and collisions (either between test particles and the planet, or between test particles and the star) constitute a more likely outcome. Although XO-3b is on a 3-day orbit, its high mass drives its Safronov number close to unity.

\begin{figure}[t]
\centering
\includegraphics[width=1\columnwidth]{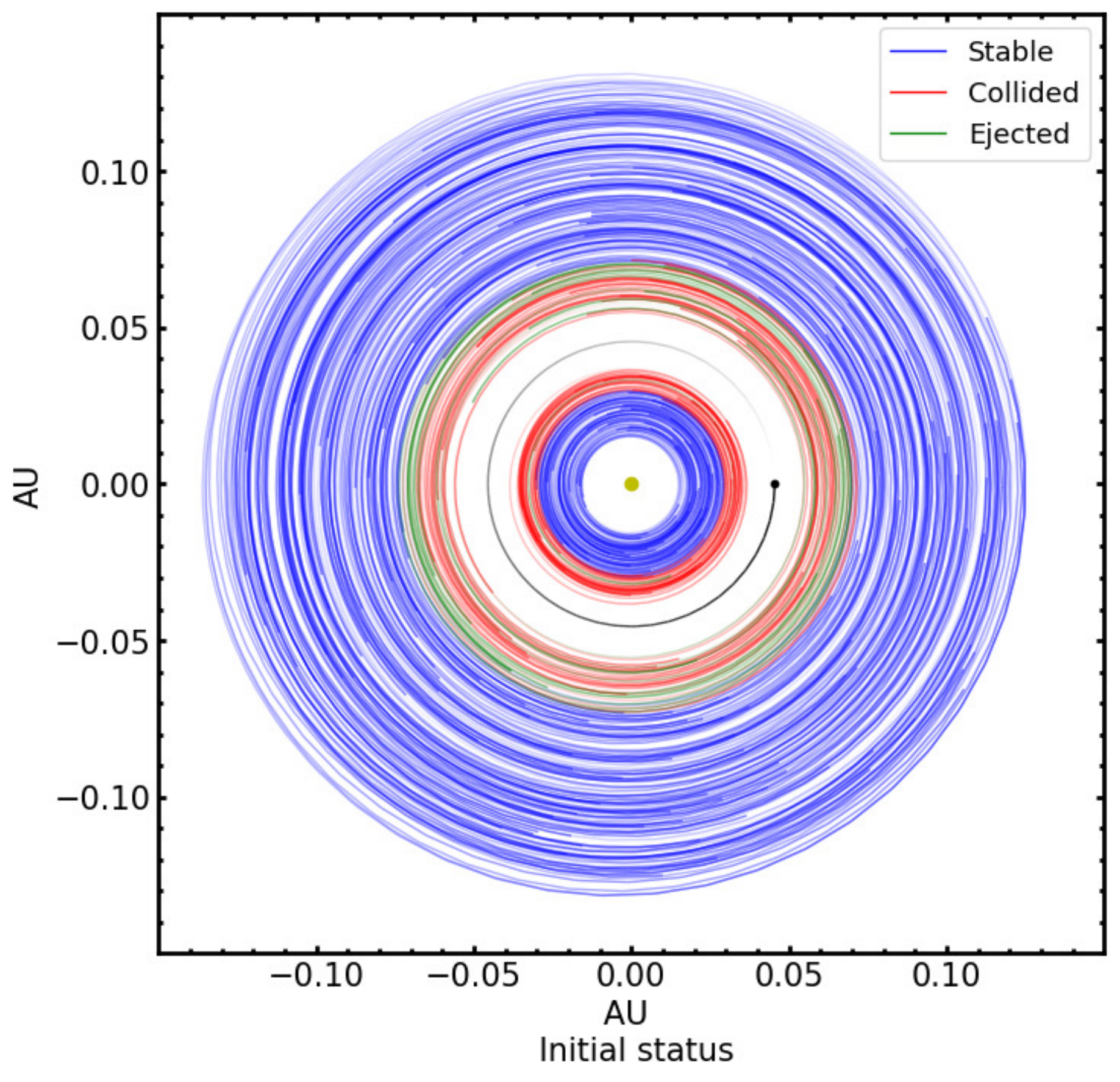}
\caption{A top-down view of the XO-3 system. Initial orbits of 300 test particles are evenly distributed over $1.3-7.5$ mutual Hill radii with XO-3b. Orbits remaining after a $10^6$ year integration are shown in blue. All orbits lying within $3.5$ mutual Hill radii of XO-3b have been destabilized. Among the unstable particles, $80\%$ collide with XO-3 (red), and, as a consequence of XO-3b's large mass, $20\%$ are ejected (green).}
\label{f6}
\end{figure}

We explored the dynamical behavior of test particles on both interior and exterior orbits between $1.3$ to $7.5$ mutual Hill radii \citep{Gladman1993} with XO-3b. As shown in Figure~\ref{f6}, the stability of test particles increases with increasing distance from XO-3b. All of the unstable particles initially lay within 3.5 mutual Hill radii of the planet, in concordance with the rule of thumb found by \citealt{Chambers1996}. In terms of the channels for instability, $80\%$ collided with XO-3, while $20\%$ were ejected from system. The integrations indicate that XO-3 is approaching (but has not reached) a regime in which small bodies in the vicinity of the planet are predominantly ejected from the system rather than incorporated into the planet. This suggests the existence of a cross-over mass slightly above that of XO-3 where further growth in mass would be restricted, and it is consistent with the observed near-absence of such objects in the short-period planet population \citep{Marcy2000}, despite their ready detectability through either transit photometry or Doppler velocity.

\begin{figure*}[t]
\centering
\includegraphics[width=1\textwidth]{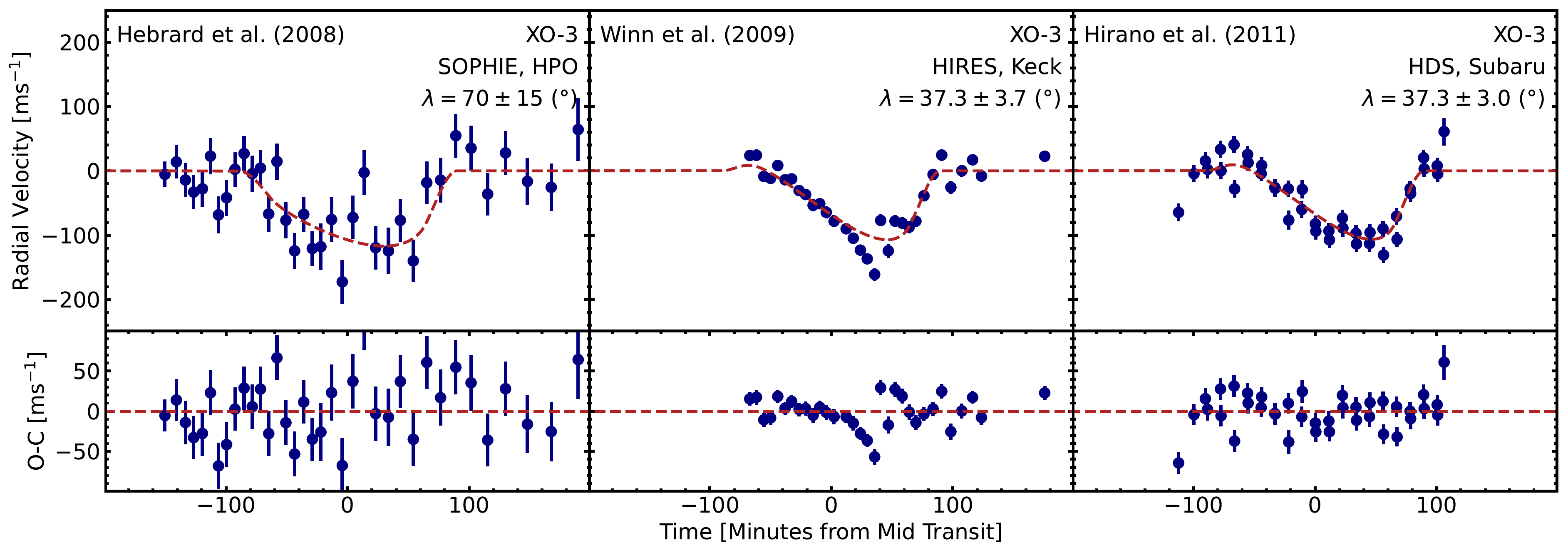}
\caption{Spectroscopic radial velocities of XO-3 from \cite{Hebrard2008} (left), \cite{Winn2009} (center), and \cite{Hirano2011} (right) as a function of orbital phase (minutes from mid-transit), coupled with the best-fitting R-M models (red dashed line). The RMS of the residuals from the three works, from left to right, is 40 $\rm m\,s^{-1}$, 18 $\rm m\,s^{-1}$, and 23 $\rm m\,s^{-1}$.}
\label{f5}
\end{figure*}

XO-3b is also known as the first exoplanet measured to have a large spin-orbit misalignment ($70.0\degree \pm 15.0\degree$, \citealt{Hebrard2008}). Follow-up studies on the stellar obliquity of the XO-3 system based on radial velocities collected with Keck/HIRES and Subaru/HDS found a smaller angle ($37.3\degree \pm 3.7\degree$, \citealt{Winn2009}; and $37.3\degree \pm 3.0\degree$, \citealt{Hirano2011}). Although there are indications of systematic effects in previous datasets (see \citealt{Hirano2011} for detailed discussion), the discrepancy could also be caused by true astrophysical reasons that warrant investigation.

\textbf{Astrophysical Origin}. The true angle between the orbital angular momentum vector of XO-3b and its stellar spin vector can be determined from three independent angles: The sky-projected spin-orbit angle ($\lambda$), which can be measured through the Rossiter-McLaughlin effect; the orbital angular momentum vector of the transiting planet along the line of sight (e.g., transit inclination, $i$), which can be determined by modelling the transit light curve; and the stellar spin vector of the star along the line of sight ($i_{*}$), which can be estimated by measuring $V_{*}\sin{i_{*}}$ from spectroscopic observation and $V_{*}$ by analyzing periodic photometric variations in the light curve caused by the stellar spots.

If the discrepancy between previous $\lambda$ measurements is truly caused by a change in the spin-orbit angle of XO-3 system, then it is statistically improbable that it only changed within the projected plane (perpendicular to the line of sight, e.g., $\lambda$). In the following analyses, we examine whether the stellar spin vector of XO-3 along the line of sight ($i_*$) or the orbital angular momentum vector of XO-3b along the line of sight (e.g., transit inclination, $i$) also changed over time. We do not find strong evidence in support of these changes.

\underline{Stellar Spin Vector along the line of sight  ($i_*$)}. Internal gravity waves within hot (and thus massive) stars \citep{Rogers2012} can induce time-dependent variations in the direction of the stellar surface spin. Given that XO-3 is a fairly hot and fairly massive star, we tested whether this type of stellar polar wander occurs in this system.

The combination of estimates of $V_* \sin{i_*}$ (from spectroscopic measurements) and $V_*$ (from light curve measurements) can provide the stellar spin vector of a star along the line of sight ($i_*$). Although the spectroscopic rotation velocity obtained from \cite{Johns-Krull2008} ($V_{\rm *before}\sin{i_{\rm *before}}=18.54 \pm 0.17 \rm \ km \ s^{-1}$) agrees with the average of \cite{Winn2009}, \cite{Hirano2011}, \cite{Torres2012}, and \cite{Brewer2016} ($V_{\rm *after}\sin{i_{\rm *after}}=17.9 \pm 0.5  \rm \ km \ s^{-1} $) within $1.2\sigma$, we cannot constrain the possible variation of the stellar spin vector ($\Delta i_*$) that might occur, since XO-3 did not leave a sufficient signal (which would typically manifest as periodic photometric variations from star spots) in its light curve to measure its true rotation velocity ($V_*$).

However, the difference between the stellar spin vector of XO-3 along the line of sight before \cite{Hebrard2008} and after \cite{Winn2009}  ($\Delta i_*=i_{\rm *after}-i_{\rm *before}=\arcsin(\frac{V_{\rm *after }\sin{i_{\rm *after}}}{V_*})-\arcsin(\frac{V_{\rm *before }\sin{i_{\rm *before}}}{V_*})$) only depends on the true rotational velocity of the star $V_*$, if we assume that the rotational velocity of the star, $V_*$, does not change over time, this means that the rotational velocity before \cite{Hebrard2008} and after \cite{Winn2009} $V_{\rm *before}=V_{\rm *after}=V_*$. The \textit{maximum} possible variation is then $\Delta i_*= 15.26\degree \pm5.82\degree $, when $V_{*} =\mathrm{max}(V_{\rm *}\sin{i_{\rm * \ before}}, V_{\rm *}\sin{i_{\rm * \ after}} ) = V_{\rm * \ before}= 18.54 \pm 0.17 \rm \ km \ s^{-1}$. This agrees with no change (e.g., $\Delta i_*=0^\circ$) within $2.6\sigma$. This is the \textit{maximum} change that is consistent with current data, since here we assume the minimum possible value of $V_*$ corresponds to the largest possible $\Delta i_*$.

\begin{figure}[t]
\centering
\includegraphics[width=1\columnwidth]{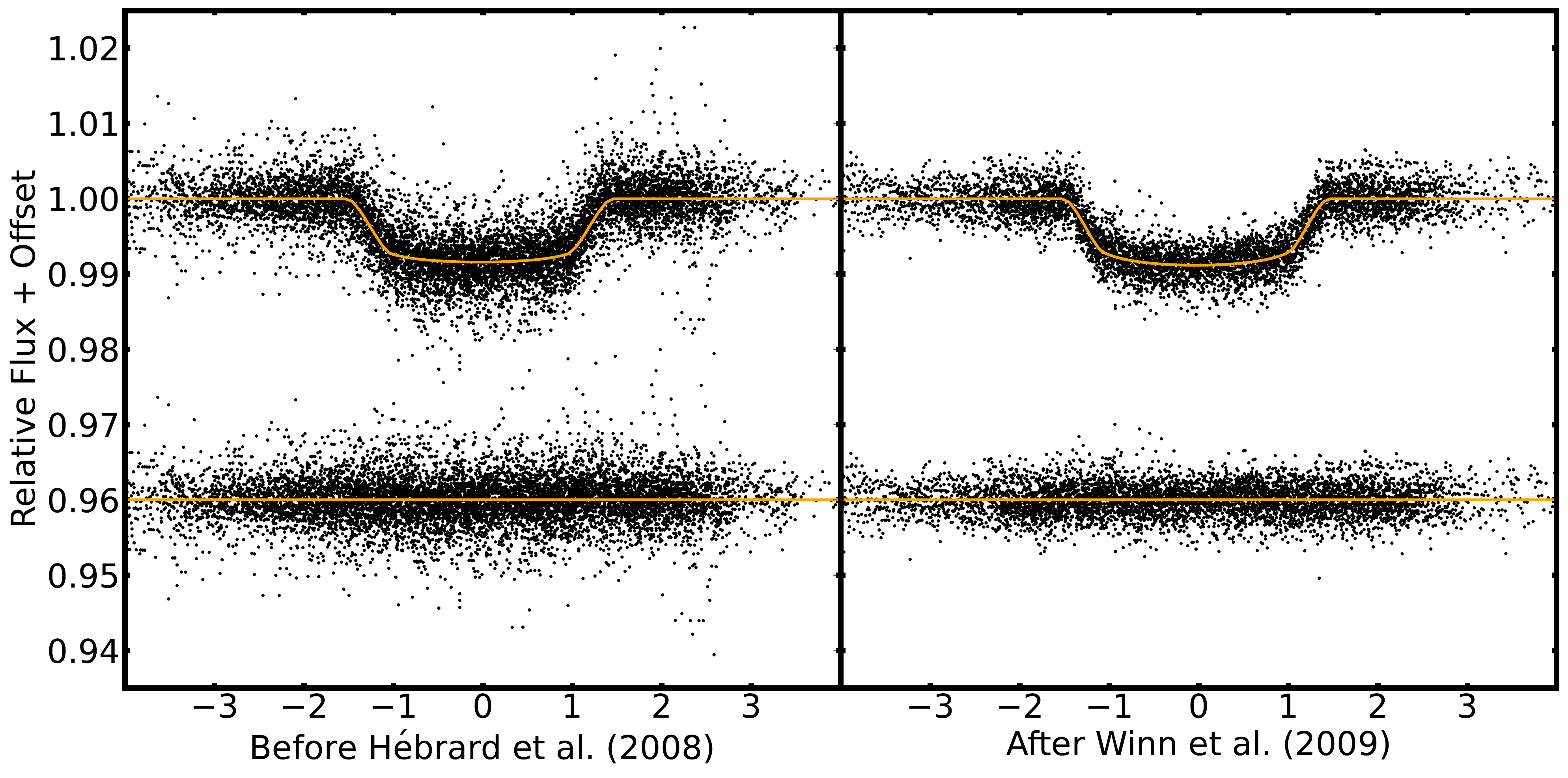}
\caption{Left: Phased transit light curves before the R-M measurement conducted by \citet{Hebrard2008}. Right: Phased transit light curves after the R-M measurement conducted by \citet{Winn2009}. The best-fitting model for each dataset is overplotted as a solid orange line. The two transit profiles are in good agreement.
}
\label{f7}
\end{figure}

\underline{Orbital Angular Momentum Vector}. Precession of nodes due to an additional perturber could also cause the temporal variation of the direction of the orbital angular momentum vector \citep{Innanen1997}. Precession can manifest as temporal variation of the transit profile, with evolution occurring in the impact parameter ($b$), the transit inclinations ($i$), the transit duration ($T_{14}$), and the transit depth ($\delta$). We thus separately fit the transit and radial velocity data before the R-M measurement conducted by \citet{Hebrard2008}, and after the R-M measurement conducted by \citet{Winn2009}. As shown in Table~\ref{tab:XO3beforeafter} and Figure~\ref{f7}, 
we found that the transit inclinations derived from the data before \citet{Hebrard2008}) and after  \citet{Winn2009} are in excellent agreement with each other with only $\Delta i=0.59^\circ\pm0.5^\circ$ difference, which means that the orbital angular momentum vector of XO-3b along the line of sight agrees with no change within $1.2\sigma$.

In conclusion, we found no evidence for temporal changes of either the stellar spin vector or the orbital angular momentum vector of XO-3b along the line of sight.

\textbf{Systematic noise}. \citet{Winn2009} and \citet{Hirano2011} suspected that the discrepancy in multiple R-M measurements for XO-3b is most likely due to systematic errors in the datasets \citep{Hebrard2008}, which will lead to underestimated uncertainties in $\lambda$.

We bin the residuals of each of three R-M measurements \citep{Hebrard2008, Winn2009, Hirano2011} into bin sizes $N=1-15$ and evaluate the RMS of the data. We found that the RMS of the time-binned residuals for all three datasets decreases more slowly than $N^{1/2}$ ($N^{1/3}=$ for \citealt{Hebrard2008}, $N^{1/5}=$ for \citealt{Winn2009}, $N^{1/3}=$ for \citealt{Hirano2011}), suggesting that strong correlated noise is present.

This is also clear in the residual panels of Figure~\ref{f5}. In this figure, we took the data directly from \citet{Hebrard2008}, \citet{Winn2009}, and \citet{Hirano2011}, and we subtracted away the baseline RV trend from each dataset using the parameters reported by each respective paper. We then plotted the reported model (using the previous authors' results) in Figure~\ref{f5}, showing the residual amplitudes below.

The data from all three datasets \citep{Hebrard2008, Winn2009, Hirano2011} suffers from strong systematic errors: that is, the residuals show clear structure that remains after the best-fitting model has been subtracted from the data. This is not surprising for a hot and massive star like XO-3, and it leads to an underestimate of uncertainties in $\lambda$.

Although current observations still allow for quite a bit of temporal variation of the spin-orbit angle of XO-3 system, we have demonstrated that the disagreement between sky-projected spin-orbit angles measured from previous studies very likely results from the underestimate of the uncertainties of $\lambda$ due to the presence of systematic noise in the data, with no requirement for a true temporal variation in the spin-orbit angle. We also find no strong evidence that two other angles have changed.

The origin and evolution of spin-orbit misalignments remains one of the most interesting unsolved problems posed by the observed properties of the exoplanets \citep{Winn2015, Albrecht2021}. A definitive assessment would be easier to formulate if measurements of spin-orbit angles could be made for a variety of system types. The R-M effect, however, is much more easily measured when transits are frequent and deep. Therefore, while R-M observations of small planets and/or long-period planets play a critical role in understanding the origin of spin-orbit misalignment \citep{Albrecht2013, Rice2021, Wang2018b,Wang2021,WangX2021b, Zhou2018}, they are difficult to successfully carry out. 

Although the R-M effect was first established through the observation of an eclipsing binary more than a century ago \citep{Schlesinger1910}, such measurements of low-mass eclipsing binaries are surprisingly rare (see BANANA Project, \citealt{Albrecht2007}). NASA's \textit{TESS} mission has been steadily discovering suitable targets (e.g., \citealt{Huang2018, Wang2019, Jones2019, Canas2019, Gunther2019}) and will detect a large number of low-mass eclipsing binaries orbiting bright stars that are suitable for R-M follow-up. This new population may shed light on not only planet formation, but also on the genesis of low-mass stars.

\acknowledgements
K.W. thanks Yale College for supporting this work through the First-Year Research Fellowship in the Sciences and the Edward A Bouchet Fellowship. S.W. thanks the Heising-Simons Foundation for their generous support as a 51 Pegasi b fellow. J.A.B. thanks MIT’s Kavli Institute for its support as a Torres postdoctoral fellow. M.R. is supported by the National Science Foundation Graduate Research Fellowship Program under Grant Number DGE-1752134. The research was carried out in part at the Jet Propulsion Laboratory, California Institute of Technology, under a contract with the National Aeronautics and Space Administration (80NM0018D0004). This work is supported by Astronomical Big Data Joint Research Center, co-founded by National Astronomical Observatories, Chinese Academy of Sciences and Alibaba Cloud. We thank Jason Eastman, John Michael Brewer, and Pia Cortes-Zuleta for useful discussions.


\end{document}